\documentclass[twocolumn]{aastex7}

\usepackage{amsmath}
\usepackage{longtable}
\usepackage{changepage}
\usepackage{rotating}
\usepackage{silence}
\usepackage{graphicx}
\usepackage{subcaption}
\usepackage{float}
\providecommand{\sw}[1]{\texttt{#1}}

\newcommand{\thisgrb}{GRB~230204B}

\newcommand{\vbdone}{}

\begin{document}

\title{GRB 230204B: GIT Discovery of a Fast Fading Afterglow Associated with an Energetic Gamma-Ray Burst from a Massive Star Progenitor}


\author[0000-0002-7942-8477]{Vishwajeet Swain}
\email{vishwajeet.s@iitb.ac.in}
\affiliation{Department of Physics, Indian Institute of Technology Bombay, Powai, 400076, India}

\author[0000-0002-6112-7609]{Varun Bhalerao}
\email{varunb@iitb.ac.in}
\affiliation{Department of Physics, Indian Institute of Technology Bombay, Powai, 400076, India}

\author[0000-0003-0871-4641]{Harsh Kumar}
\email{harsh.kumar@cfa.harvard.edu}
\affiliation{Department of Physics, Indian Institute of Technology Bombay, Powai, 400076, India} \affiliation{Center for Astrophysics \textbar{} Harvard \& Smithsonian, 60 Garden Street, Cambridge, MA 02138-1516, USA}

\author[0009-0004-9984-4138]{Mehul Goyal}
\email{mehulgoyal.iitb@gmail.com}
\affiliation{Department of Physics, Indian Institute of Technology Bombay, Powai, 400076, India}

\author[0000-0003-2265-0381]{Ankur Ghosh}
\email{ghosh.ankur1994@gmail.com}
\affiliation{Centre for Astro-Particle Physics (CAPP) and Department of Physics, University of Johannesburg, PO Box 524, Auckland Park 2006, South Africa}

\author[0009-0002-7897-6110]{Utkarsh Pathak}
\email{utkarshpathak.07@gmail.com}
\affiliation{Department of Physics, Indian Institute of Technology Bombay, Powai, 400076, India}

\author[0000-0002-0844-6563]{Poonam Chandra}
\email{vishwajeet.s@iitb.ac.in}
\affiliation{National Radio Astronomy Observatory, Charlottesville VA 22903, USA}

\author[0000-0002-2184-6430]{Tomás Ahumada}
\email{tahumada@astro.umd.edu}
\affiliation{Cahill Center for Astrophysics, California Institute of Technology, MC 249-17, 1216 E California Boulevard, Pasadena, CA 91125, USA}

\author[0000-0003-3533-7183]{G. C. Anupama}
\email{gca@iiap.res.in}
\affiliation{Indian Institute of Astrophysics, II Block Koramangala, Bengaluru 560034, India}

\author[0000-0002-6657-9022]{S. Bala}
\email{sumanbala2210@gmail.com}
\affiliation{Science and Technology Institute, Universities Space Research Association, Huntsville, AL 35805, USA}

\author[0000-0002-3927-5402]{Sudhanshu Barway}
\email{sudhanshu.barway@iiap.res.in}
\affiliation{Indian Institute of Astrophysics, II Block Koramangala, Bengaluru 560034, India}

\author[0000-0002-7777-216X]{Joshua S. Bloom}\email{joshbloom@berkeley.edu}
\affiliation{Department of Astronomy, University of California, Berkeley, CA 94720-3411, USA}

\author[0000-0001-9868-9042]{Dimple}
\email{d.dimple@bham.ac.uk}
\affiliation{School of Physics and Astronomy, University of Birmingham, Edgbaston, Birmingham, B15 2TT, UK, Institute for Gravitational Wave Astronomy, University of Birmingham, Birmingham, B15 2TT, UK.}

\author[0000-0003-2758-159X]{Viraj Karambelkar}
\email{}
\affiliation{Cahill Center for Astrophysics, California Institute of Technology, MC 249-17, 1216 E California Boulevard, Pasadena, CA, 91125, USA}

\author[0000-0002-5619-4938]{Mansi Kasliwal}
\email{mansi@astro.caltech.edu}
\affiliation{Cahill Center for Astrophysics, California Institute of Technology, MC 249-17, 1216 E California Boulevard, Pasadena, CA, 91125, USA}

\author[0000-0003-1637-267X]{Kuntal Misra}
\email{kuntal@aries.res.in}
\affiliation{Aryabhatta Research Institute of Observational Sciences, Manora Peak, Nainital 263129, India}

\author[0000-0003-1227-3738]{Josiah Purdum}\email{jpurdum@caltech.edu}
\affiliation{Caltech Optical Observatories, California Institute of Technology, Pasadena, CA  91125}

\author[0000-0001-6332-1723]{Divita Saraogi}
\email{divitadsaraogi@gmail.com}
\affiliation{Department of Physics, Indian Institute of Technology Bombay, Powai, 400076, India}

\author[0000-0003-1546-6615]{Jesper Sollerman}\email{jesper@astro.su.se}
\affiliation{The Oskar Klein Centre, Department of Astronomy, Stockholm University, AlbaNova, SE-10691 Stockholm, Sweden}

\author[0009-0005-8230-030X]{Aswin Suresh}
\email{aswinsuresh2029@u.northwestern.edu}
\affiliation{Department of Physics and Astronomy, Northwestern University, Evanston, IL 60208, USA} \affiliation{Center for Interdisciplinary Exploration and Research in Astronomy, Northwestern University, 1800 Sherman Avenue, Evanston, IL 60201, USA}

\author[0000-0001-9276-1891]{Stéfan van der Walt}
\email{stefan@mentat.za.net}
\affiliation{Berkeley Institute for Data
Science, University of California Berkeley, Berkeley, CA 94720, USA}

\author[0000-0003-3630-9440]{Gaurav Waratkar}
\email{gauravwaratkar@iitb.ac.in}
\affiliation{Cahill Center for Astrophysics, California Institute of Technology, MC 249-17, 1216 E California Boulevard, Pasadena, CA, 91125, USA} \affiliation{Department of Physics, Indian Institute of Technology Bombay, Powai, 400076, India}



\begin{abstract}
We present a comprehensive multiwavelength study of a bright gamma-ray burst GRB~230204B, analyzing both prompt and afterglow emissions. This GRB is highly energetic, with an isotropic equivalent energy emission $E_{\mathrm{iso}} \sim 2.2 \times 10^{54}\ \mathrm{erg}$ released during the prompt emission. The GROWTH-India Telescope discovered a bright afterglow ($m_r=15.55$) that fades rapidly ($\propto t^{-1.82}$). The prompt emission shows a strong thermal photospheric emission along with a nonthermal high-energy component. We explore the evolution of these components and find them to be consistent with the theoretical expectations of the fireball model. Afterglow modeling reveals an energetic jet ($E_{\gamma} \gtrsim 10^{52}$~erg) expanding into a wind-type medium viewed nearly on-axis, suggesting a massive star progenitor with strong winds. We also explore correlations between the prompt emission and afterglow that may help \added{to place GRB~230204B within the broader context of the long GRB population.}

\end{abstract}


\keywords{
\uat{Gamma-ray bursts}{629}
\uat{Burst astrophysics}{187}
\uat{Relativistic jets}{1390}
\uat{Wolf-Rayet stars}{1806}
}


\section{Introduction}\label{sec:introduction}
\vbdone Long gamma-ray bursts (LGRBs) are highly energetic explosions with strong relativistic jets, associated with the death of massive stars \citep{1999ApJ...524..262M, 2004RvMP...76.1143P, Woosley_2006}. Observations indicate that the collimation-corrected gamma-ray energy release from gamma-ray bursts (GRBs) is typically on the order of $10^{51}~\mathrm{erg}$, with isotropic equivalent energy ($E_{\mathrm{iso}}$) spanning a broad range from $10^{48}$ to $10^{55}~\mathrm{erg}$ \citep{2004RvMP...76.1143P, 2015PhR...561....1K}. GRBs with $E_{\mathrm{iso}} > 10^{54}~\mathrm{erg}$ are commonly classified as energetic GRBs \citep{2023ApJ...949L...4L}. The collapse of rapidly rotating Wolf--Rayet (W-R) stars is considered a plausible progenitor channel for these highly energetic events \citep{Woosley_2006}.

\vbdone The prompt emission of GRBs is usually nonthermal in nature, and the time-integrated spectrum is typically modeled by the empirical Band function \citep{Band:1993, Massaro_2010, 2012MNRAS.424.2821V}. However, this model lacks a direct physical interpretation. In time-resolved spectroscopy using smaller time bins, the spectra can sometimes still be fit with a Band function, but in other cases, they can be fit with physically motivated models such as blackbody (BB), cutoff power law (CPL), or combinations like power law plus blackbody (PL+BB), Band+BB, etc. \citep{Ryde_2010, 2011ApJ...730..141Z, 2017ExA....44..113B}.

\vbdone The standard fireball model predicts the presence of photospheric emission, along with nonthermal emission from internal shocks \citep{1978MNRAS.183..359C, 1994ApJ...430L..93R}. The spectral shape of photospheric emission is typically a BB, but under certain physical conditions, it can broaden and resemble the Band function \citep{2011ApJ...730..141Z, 10.1093/mnras/stt863}. Some GRBs, like GRB~110721A \citep{10.1093/mnras/stt863}, are well described by BB+Band models, where the Band component is interpreted as nonthermal emission.

\vbdone The spectral properties can thus be used to interpret the nature of the emission components. For instance, GRB~090902B \citep{Ryde_2010}, with $E_{\mathrm{iso}} \sim 10^{52}~\mathrm{erg}$, is well fit by the PL+BB model, indicating signatures of photospheric emission. In such models, the power-law (PL) component is attributed to nonthermal emission, while the BB component originates from the photosphere of the expanding relativistic jet \citep{Ryde_2010}. The signature of the BB component in the GRB prompt spectra could indicate that the jet is matter dominated and is not highly magnetized \citep{Veres_2012, 10.1093/mnras/stt863}. On the other hand, the energetic GRB~080916C displays a featureless spectrum well fit by a Band function alone and shows characteristics consistent with a Poynting flux-dominated jet with a high magnetic field \citep{2011ApJ...730..141Z}.

\vbdone Overall, the composition of GRB jets --- whether they are matter dominated, Poynting flux dominated, or exhibit a hybrid nature --- remains an open question. The role of magnetization near or above the photosphere is also still not fully understood.

\vbdone After the release of gamma-ray energy during the prompt emission phase, the fireball ejecta, moving at relativistic speeds, interact with the circumburst medium (CBM) and produce multiwavelength afterglow emission \citep{1998ApJ...497L..17S, 2003ApJ...589L..69L, Peer_2024}. More than 500 GRB afterglows have been detected and studied since the first X-ray and optical discovery in 1997 February \citep{10.1093/mnras/stae1484}. Over the past few years, wide-field time-domain surveys have significantly advanced our ability to discover and characterize optical counterparts of high-energy transients. In particular, the Zwicky Transient Facility \citep[ZTF;][]{Bellm2019PASP, Dekany2020PASP, Graham2019PASP, Masci2019PASP} has played a leading role in the systematic discovery and follow-up of GRB afterglows and fast X-ray transients, owing to its high cadence and wide sky coverage. LGRBs are usually associated with massive star progenitors, whose environments are expected to be shaped by stellar winds released prior to the burst \citep{Chevalier_2000}. Although some GRBs are consistent with the expected wind-type environments, many others are better fitted by a uniform density medium \citep{2001ApJ...560L..49P, Li2003, Wang_2015}. Other works suggest that a uniform density region surrounding a massive star can result from a wind termination shock, which becomes possible if the mass-loss rate is reduced to $10^{-6}~M_{\odot}~\mathrm{yr}^{-1}$ due to the low metallicity of the progenitor star \citep{Wijers_1999,Li2003}. 

\vbdone For some sources, it is difficult to distinguish between an ISM and a wind environment using data available only a few days postburst \citep{2003ApJ...589L..69L, 2001ApJ...560L..49P}.  A wind-type CBM is typically identified using radio data, where the light curve is expected to remain flat for days. Another possible signature of the wind-type medium is a steep decay in the optical light curve. However, this can also be caused by a highly collimated jet \citep{Sari_1999, 2001ApJ...560L..49P}.

\added{In this paper, we report the discovery of the optical afterglow of \thisgrb\ and its rapid fading behavior \citep{2023GCN.33269....1S} using the GROWTH-India Telescope (GIT; \citealt{2022AJ....164...90K}), together with a comprehensive set of multiwavelength observations obtained from several facilities. In particular, our dataset also includes late-time radio detections that reveal a long-lived radio afterglow component. Details of the observations and data reduction procedures are presented in Section~\ref{sec:observations} and Appendix~\ref{appendixData}, respectively.

In this work, we perform a detailed time-integrated and time-resolved spectral analysis of the prompt emission using the PL+BB model and compare the results from the Band and CPL models. We particularly characterize the evolution of the thermal (photospheric) and nonthermal components to directly constrain key jet properties, including the bulk Lorentz factor, photospheric radius, and the base radius of the outflow. We also probe the possible origin of the thermal and nonthermal components. The detailed prompt analysis is presented in Section~\ref{sec:prompt_analysis}.

Next, we analyze the evolution of the afterglow by combining the rapidly fading optical afterglow with late-time radio detections. Based on these observational constraints, we have ruled out a uniform density environment and instead modeled the afterglow in a wind-type CBM. Our best-fit model successfully explains both the observed steep decay in optical brightness and the nearly constant radio flux. A detailed description of the afterglow modeling can be found in Section~\ref{sec:afterglow}. Finally, we place \thisgrb\ within the population of long GRBs by exploring empirical correlations between prompt emission and afterglow properties in Section~\ref{sec:discussion}. We also discuss the physical factors that contribute to the rapid afterglow decay and the implications of the progenitor system.}

\section{Observations}\label{sec:observations}
\vbdone \thisgrb\ was first detected by the \emph{MAXI-GSC} nova alert system at trigger time ($T_0$), 21:47:51 UT on 2023 February 4 \citep{2023GCN.33265....1S}. A bright source was located at the position of R.A. (J2000): 13h 10m 19s and decl. (J2000): $-$21d 45m 07s, with an elliptical error region long and short radii of $0.12^\circ$ and $0.1^\circ$ at the statistical $90\%$ confidence level. 
The burst was independently detected by multiple high-energy instruments, including \emph{Fermi}-GBM \citep{2023GCN.33288....1P}, \emph{AstroSat}-CZTI \citep{2023GCN.33268....1W}, \emph{AGILE} \citep{2023GCN.33272....1C}, and \emph{GRBAlpha} \citep{2023GCN.33273....1D}. Throughout this work, we use the \emph{Fermi} trigger time $T_0=$ 2023 February 4 21:44:27.20 as the reference time.

\begin{figure}[ht]
    \centering
    \includegraphics[width=0.8\columnwidth]{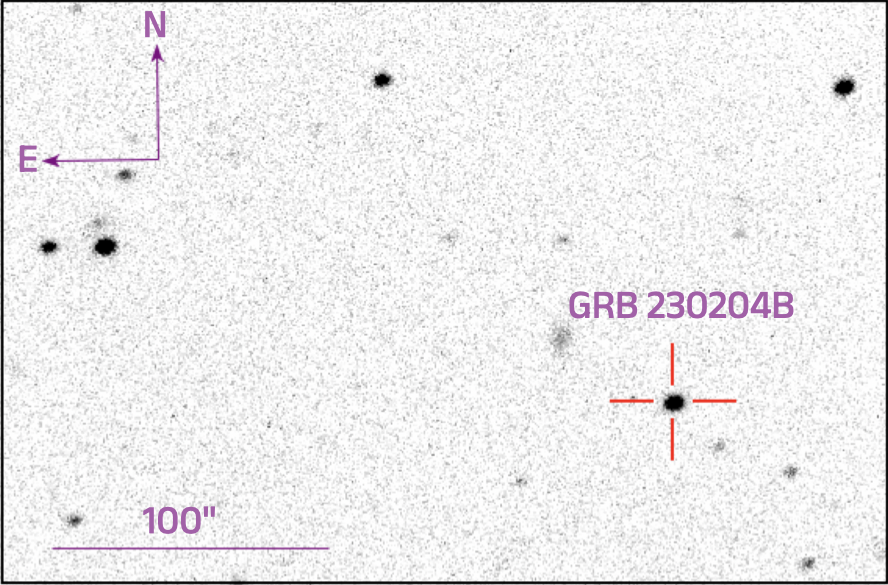}
    \caption{The optical afterglow of \thisgrb\ was discovered by GIT in the $r'$ filter at UT 2023 February 4 23:22:37.6, as reported by \citet{2023GCN.33269....1S}. The source was both bright and rapidly fading, consistent with the expected behavior of a GRB afterglow.} 
    \label{fig:GIT_image}
\end{figure}

\vbdone GIT was triggered as soon as the target became visible, $\approx$ 1.5~hr after the \emph{Fermi} GBM trigger time $T_0$, to search for the optical counterpart. GIT identified a new source located at R.A. (J2000): 13h 10m 19s and decl. (J2000): $-$21d 45m 07s, with an uncertainty of $0.67\arcsec$, as shown in Figure~\ref{fig:GIT_image}. The source was observed to be fading rapidly, confirming it as the afterglow of the GRB. Using the initial large dataset obtained in the $r$-band, GIT reported its discovery of the optical afterglow and the rapid temporal decay of the source on the General Coordinates Network \citep{2023GCN.33269....1S}. As part of the GROWTH collaboration \citep{2019PASP..131c8003K}, we triggered additional observations with several other telescopes (Appendix~\ref{appendixData}). We utilize the ZTF Fritz marshal to trigger the telescopes and store the photometric data \citep{2019JOSS....4.1247V, Coughlin_2023}. Unfortunately, due to poor weather conditions at some of the  observatories, we obtained a limited dataset. Independent Very Large Telescope (VLT)/X-shooter observations captured the GRB afterglow spectrum and determined the redshift to be $z$ = 2.142 \citep{2023GCN.33281....1S}. 

Independent observations and analyses of \thisgrb\ have been reported by \citet{Gupta_2026} (hereafter G26). We incorporate the data from G26 and compare the results where applicable. Additionally, we include publicly available photometric detections from the Asteroid Terrestrial-impact Last Alert System \citep[ATLAS,][]{2023GCN.33278....1S}, 3.6-meter DOT \citep{2023GCN.33284....1R, Gupta_2026}, VLT/X-shooter \citep{2023GCN.33281....1S}, and Australia Telescope Compact Array \citep[ATCA,][]{2023GCN.33321....1G}, as published through the GCN. All the photometric data points are presented in the supplementary material at \cite{swain_2025_17782846}. The observations and data reduction are described in Appendix \ref{appendixData}.

Throughout this work, we adopt the \texttt{Planck 2018} cosmology \citep{Planck2018} in \texttt{astropy} with $H_0 = 67.66~\mathrm{km~s^{-1}~Mpc^{-1}}$, $\Omega_m = 0.30966$, and $\added{\Omega_\Lambda = 0.68884}$. For this redshift, the corresponding luminosity distance is $D_L = 5.34 \times 10^{28}~\mathrm{cm}$ (or 17.3~Gpc).

\section{Prompt emission: analysis} \label{sec:prompt_analysis}
\begin{figure}[!thb]
    \centering
    \includegraphics[width=\columnwidth]{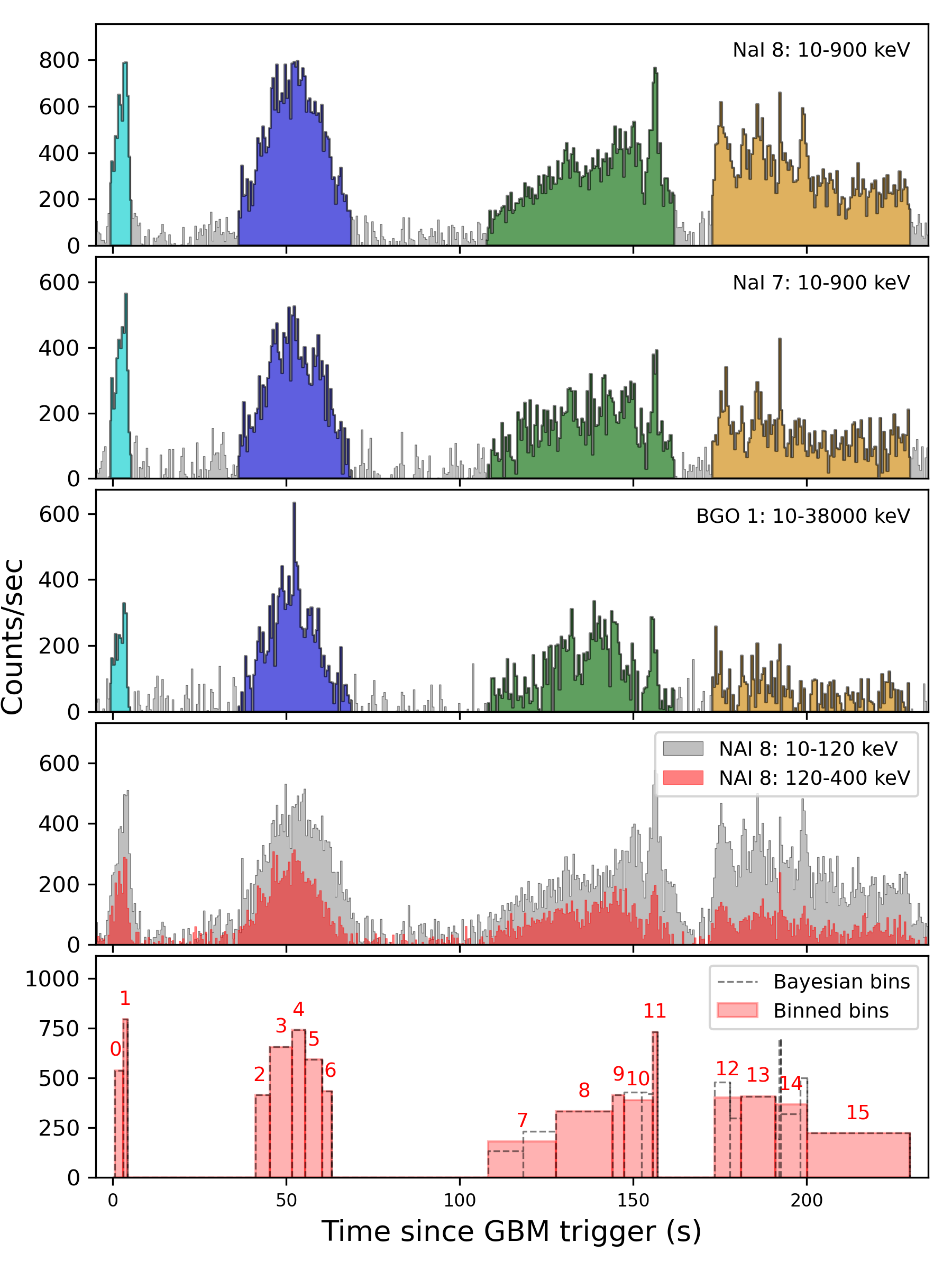}
    \caption{Multipanel light-curve analysis of \thisgrb\ using Fermi-GBM detectors. Top panels: background-subtracted light curves from NaI detectors 8 and 7 (10--900~keV), and BGO detector 1 (10--38,000~keV), color coded by episode. Middle panel: comparison of low-energy (10--120~keV) and mid-energy (120--400~keV) light curves from NaI 8, highlighting the spectral evolution during the burst. The structure reveals four distinct emission episodes, with significant variability in intensity and spectral hardness across them. Bottom panel: Bayesian blocks (dashed gray) and final binned intervals (shaded red) used for time-resolved spectral analysis, labeled with bin numbers. The binning ensures sufficient counts in high-energy channels (e.g., BGO 1) for reliable spectral fitting. The figure is plotted with the packages -- \texttt{GDT-Core} \citep{GDT-Fermi} and \texttt{ThreeML} \citep{2015arXiv150708343V}.}
    \label{fig:multi_panel}
\end{figure}

We analyzed the \emph{Fermi}-GBM prompt emission of \thisgrb\ following standard \texttt{ThreeML} procedures \citep{2015arXiv150708343V}. The time-integrated spectrum was extracted over the interval $T_0-0.81$~s to $T_0+229.96$~s. \added{The prompt emission light curve displays four distinct episodes, suggestive of intermittent central engine activity.} The time span of each episode is listed in Table~\ref{tab:prompt_model_fit}. We used the energy ranges 10--900~keV for NaI detectors and 250~keV--38~MeV for BGO detectors. The background-subtracted light curves are shown in Figure~\ref{fig:multi_panel}.

To characterize the prompt emission, we fitted both the time-integrated spectrum and individual emission episodes using commonly adopted GRB spectral models: Band, CPL, and PL+BB. Model comparison was performed using the Akaike and Bayesian information criteria (AIC and BIC), with parameter estimates summarized in Table~\ref{tab:prompt_model_fit}. While the Band and CPL models provide statistically acceptable fits, adding a BB component to the Band and CPL models does not significantly improve the fits, and the BB parameters in these cases are poorly constrained. Similar results have been reported in G26; therefore, we exclude these combinations from further discussion in this work.

The time-integrated spectrum is best described by the PL+BB model, yielding a BB temperature of $kT=65.4\pm3.0$~keV and a PL index of $\Gamma_{\rm PL} = -1.49 \pm 0.03$. The parameters obtained from the Band model fit show a low-energy spectral index of $\alpha_{\rm p} = -0.85^{+0.05}_{-0.03}$, a high-energy spectral index of $\beta_{\rm p} = -3.39^{+0.91}_{-0.15}$, and a peak energy of $E_{\rm p} = 560^{+48}_{-63}$~keV. For comparison, we also fit the spectrum with a CPL model; the resulting values of $\alpha_{\rm p}$ and $E_{\rm p}$ are consistent with those obtained from the Band model.

In the analysis of prompt emission episodes, the two brightest episodes (Episodes 2 and 3) are best described by the Band function, while Episodes 1 and 4 are better fitted by the PL+BB model. In all episodes, the Band model provides a better fit than the CPL model, with consistent values of the low-energy spectral index $\alpha_{\rm p}$ and the peak energy $E_{\rm p}$. The low-energy spectral index evolves from a hard value of $\alpha_{\rm p} = -0.34 \pm 0.10$ to a softer value of $\alpha_{\rm p} = -0.82 \pm 0.05$ throughout the burst. Notably, the value $\alpha_{\rm p} = -0.34 \pm 0.10$ in Episode~1 is harder than the typical canonical value of $\alpha_{\rm p} = -1$ for GRBs \citep{2011ApJ...730..141Z,2021ApJ...913...60P}. This harder spectrum suggests the presence of photospheric emission \citep{Ryde_2010,2011ApJ...730..141Z}. 

\begin{figure}[!thb]
    \centering
    \includegraphics[width=\columnwidth]{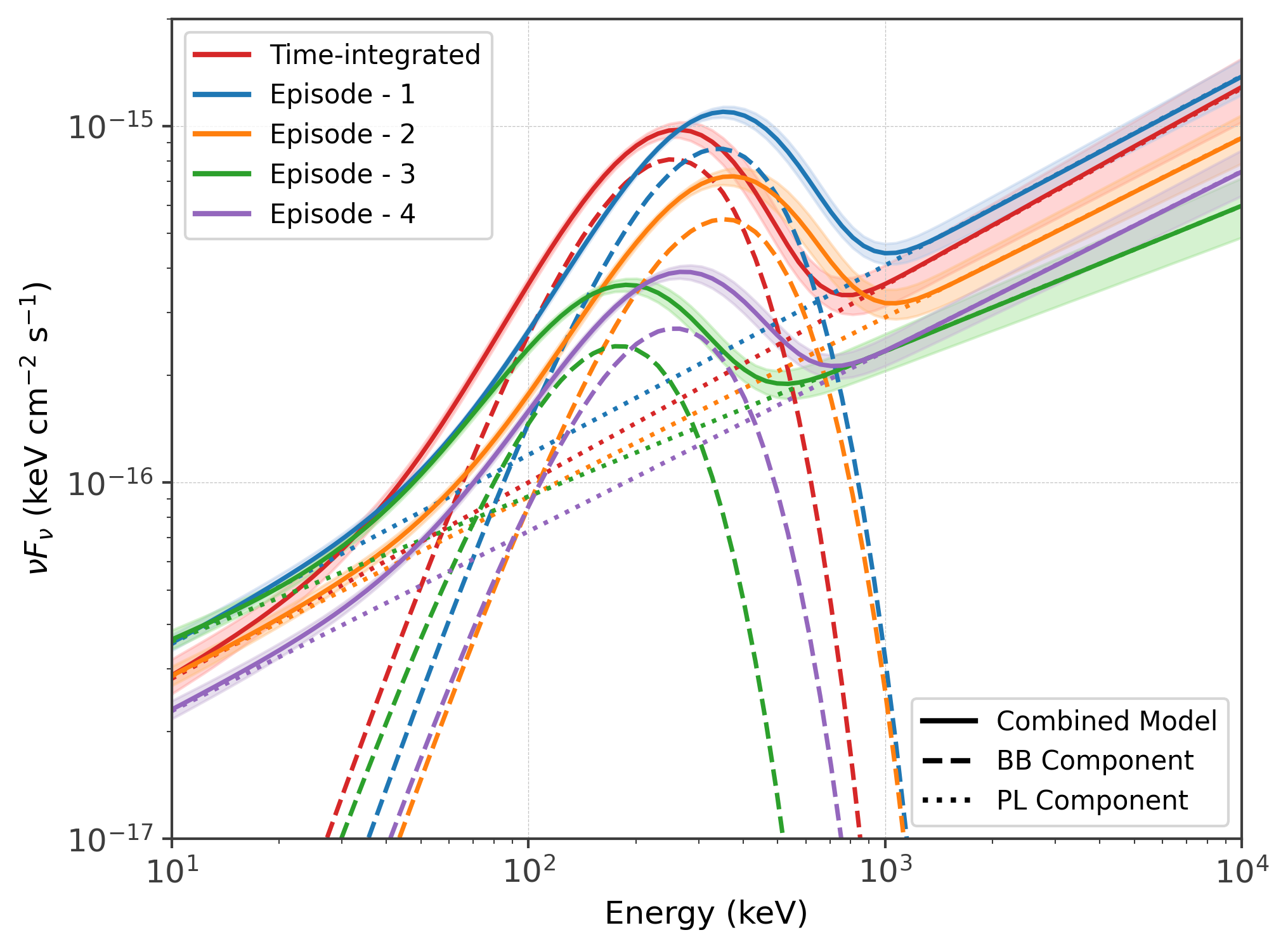}
    \caption{Time-integrated and episode-wise prompt emission spectra of \thisgrb\ modeled with a PL+BB function. Solid lines represent the best-fit combined model, while the shaded regions denote the 68\% credible intervals. The thermal and nonthermal components are shown by dashed and dotted lines, respectively. Different colors correspond to the time-integrated spectrum and individual emission episodes. The presence of a distinct thermal component, peaking at 200--500~keV across multiple epochs, provides visual evidence for photospheric emission during the prompt phase.}
    \label{fig:prompt_spec}
\end{figure}

Both the BB temperature and the peak energy exhibit a similar trend across the episodes. Episode~3 shows the hardest spectrum among all episodes, with a peak energy of $E_{\rm p} = 796 \pm 87$~keV and a BB temperature of $90.2\pm5.0$~keV. To visually illustrate the presence of the thermal component, Figure~\ref{fig:prompt_spec} displays the time-integrated and episode-wise prompt emission spectra modeled using the PL+BB function.

Comparing the fluence in the observed energy range (10--38~MeV) of each episode, we find that the first episode contributes $\sim6\%$ of the total energy output, the second contributes $\sim33\%$, the third $\sim36\%$, and the fourth $\sim25\%$. \added{The fractional contribution of the BB component is $18.7\pm1.9\%$, $18.4\pm1.6\%$, and $17.3\pm1.7\%$ for Episodes~1--3, respectively, and decreases to $11.9\pm1.8\%$ in Episode~4.} Such a decline is expected as the photosphere expands, cools, and becomes transparent, allowing nonthermal processes to dominate. The early episodes, therefore, appear consistent with a hot, weakly magnetized fireball in which photospheric emission is prominent.

\begin{table*}
    \caption{Time-averaged Spectral Fitting Parameters for time-integrated and episode-wise analysis of \thisgrb}
    \centering
    \setlength\extrarowheight{5pt}
    \begin{tabular}{l|c|c|c|c|c|c}
    \hline
        Function & Parameters & Time-Integrated & Episode 1 & Episode 2 & Episode 3 & Episode 4 \\
    \hline
    \hline
    Time range & Start time (s) & $-0.81$ & $-0.81$ & $ 36.16 $ & $ 108.22 $ & $ 171.65 $ \\
    (from $T_0$) & End time (s) & $229.96$ & $5.50$ & $69.13$ & $162.56$ & $229.96$\\
    \hline
    \hline
        & K  ($\times10^{-3}$)  & $7.77 \pm 0.36$ & $27.6 \pm 2.3$ & $14.1 \pm 0.4$ & $9.2 \pm 0.3$ & $13.3 \pm 1.1$ \\
        & $\mathrm{keV^{-1}~s^{-1}~cm^{-2}}$ & &  &  &  \\
        \textbf{Band} & $\alpha_{\rm p}$ & $-0.83^{+0.06}_{-0.04}$ & $-0.34^{+0.09}_{-0.11}$ & $-0.67^{+0.04}_{-0.04}$ & $-0.79^{+0.05}_{-0.05}$ & $-0.82^{+0.07}_{-0.05}$ \\
        & $\beta_{\rm p}$ & $-3.39^{+0.91}_{-0.15}$ & $-2.40^{+0.13}_{-0.69}$ & $-2.91^{+0.33}_{-0.31}$ & $-3.38^{+0.66}_{-0.05}$ & $-2.72^{+0.33}_{-0.46}$ \\
        & $E_{\rm p}$ (keV) & $560^{+48}_{-63}$ & $371^{+61}_{-30}$ & $690^{+53}_{-44}$ & $796^{+88}_{-86}$ & $288^{+27}_{-29}$ \\
        & AIC & 7340 & 3473 & 5311 & 5875 & 5801 \\
        & BIC & 7356 & 3462 & 5300 & 5866 & 5794 \\
    \hline
    \hline
    & K  & $0.33 \pm 0.06$ & $0.13 \pm 0.05$ & $0.31 \pm 0.05$ & $0.35 \pm 0.07$ & $0.60 \pm 0.13$ \\
    & $\mathrm{keV^{-1}~s^{-1}~cm^{-2}}$ &  & &  &  &  \\
    \textbf{CPL} & $\alpha_{\rm p}$ & $-0.81^{+0.03}_{-0.06}$  &  $-0.40^{+0.10}_{-0.10}$ & $-0.67^{+0.03}_{-0.04}$ & $-0.79^{+0.03}_{-0.05}$ & $-0.83^{+0.04}_{-0.07}$ \\
    & $E_{\rm p}$ ($\mathrm{keV}$) & $552^{+68}_{-36}$ & $424^{+50}_{-22}$  & $704^{+55}_{-37}$ & $786^{+123}_{-56}$ & $300^{+34}_{-17}$ \\
    & AIC & 7361 & 3484 & 5330 & 5895 & 5822 \\
    & BIC & 7372 & 3496 & 5341 & 5906 & 5833 \\
    \hline
    \hline
        & $K_{\rm PL}$ & $2.71 \pm 0.30$ & $2.95 \pm 0.56$ & $4.02 \pm 0.35$ & $3.46 \pm 0.39$ & $5.35 \pm 0.67$ \\
        & $\mathrm{keV^{-1}~s^{-1}~cm^{-2}}$ & & &  &  &  \\
        & $K_{\rm BB}$ ($\times10^{-6}$) &$1.19 \pm 0.18$ & $3.71 \pm 0.57$ & $1.22 \pm 0.13$ & $0.67 \pm 0.12$ & $4.49 \pm 0.73$ \\
        \textbf{PL+BB} & $\mathrm{keV^{-3}~s^{-1}~cm^{-2}}$ && &  &  &  \\
        & $\Gamma_{\rm PL}$ &$-1.49^{+0.03}_{-0.04}$& $-1.44^{+0.03}_{-0.05}$ & $-1.46^{+0.02}_{-0.02}$ & $-1.49^{+0.02}_{-0.03}$ & $-1.58^{+0.02}_{-0.04}$ \\
        & $kT$ ($\mathrm{keV}$) &$65.4^{+3.2}_{-2.9}$ & $64.9^{+3.2}_{-3.0}$ & $87.32^{+2.8}_{-3.2}$ & $90.26^{+4.8}_{-5.2}$ & $45.6^{+2.3}_{-2.0}$ \\
        & AIC & 7327 & 3468 & 5363 & 5896 & 5778 \\
        & BIC & 7342 & 3483 & 5379 & 5911 & 5793 \\
    \hline
    \end{tabular}
    \label{tab:prompt_model_fit}
\end{table*}

\subsection{Time-resolved Analysis}\label{sec:time_resolved_analysis}

\begin{figure*}[tpbh]
    \centering
    \includegraphics[width=1\textwidth]{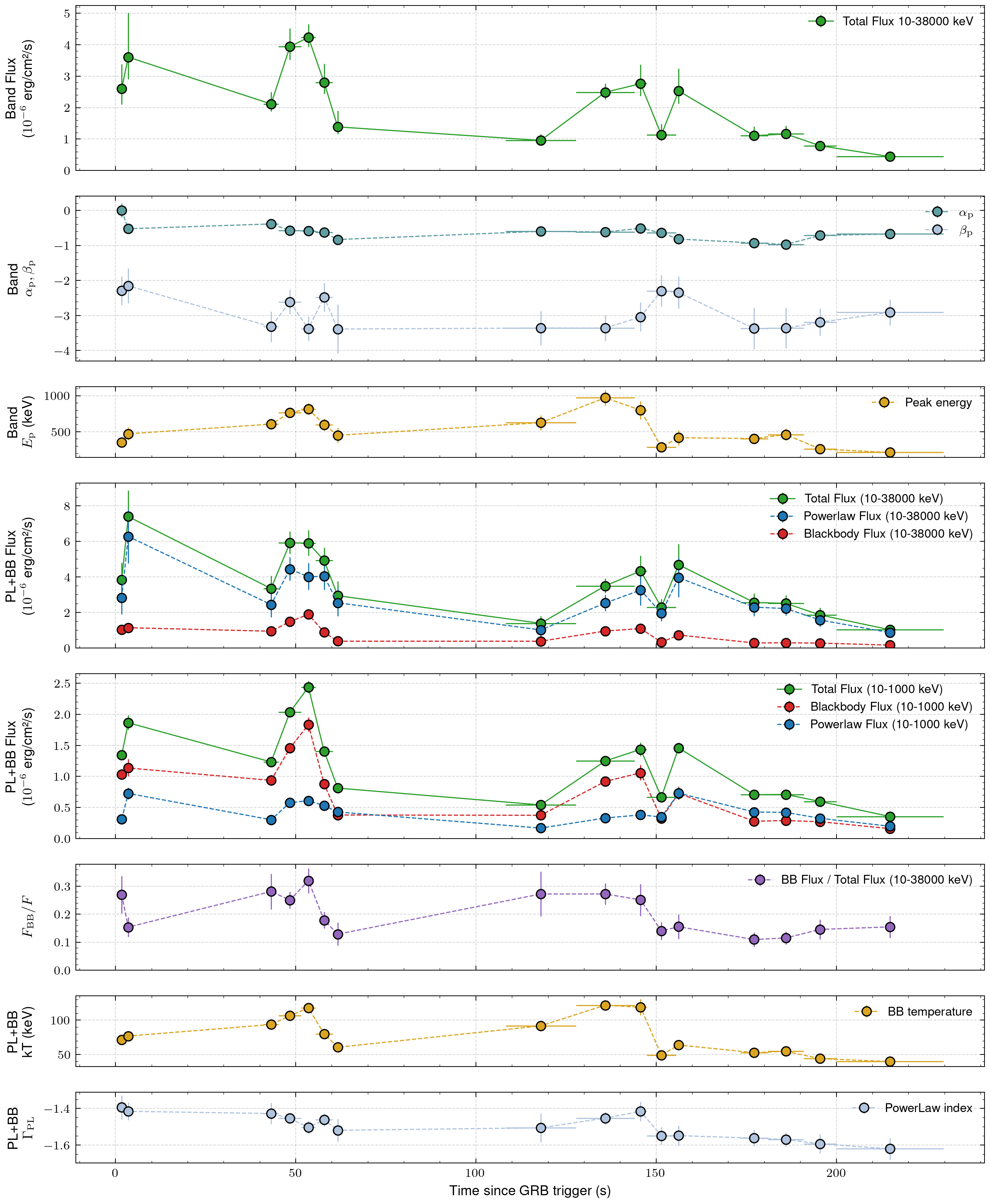}
    \caption{Time-resolved spectral evolution of \thisgrb\ using both the Band and PL+BB models.
    Top three panels: light curve and spectral evolution from the Band function fits in the 10--38,000~keV range. The panels show (i) total flux, (ii) low and high-energy photon indices ($\alpha_{\rm p}$, $\beta_{\rm p}$), and (iii) peak energy ($E_{\rm p}$).
    Bottom five panels: temporal evolution from the PL+BB model. The panels show (iv) total flux along with separate contributions from the PL and BB components in 10--38,000~keV, (v) their fluxes in 10--1000~keV, (vi) the ratio of BB to total flux, (vii) BB temperature ($kT$), and (viii) the PL photon index ($\Gamma_{\rm PL}$).}
    \label{fig:tr_flux_evolution}
\end{figure*}

For the time-resolved analysis, we used the brightest NaI detector (NaI~8) and divided the light curve using Bayesian blocks with a false-alarm probability threshold of $P=0.01$ \citep{2013ApJ...764..167S}. To ensure sufficient signal-to-noise in the high-energy BGO~1 spectra, consecutive blocks were merged until a minimum significance of 5$\sigma$ was achieved in each bin. Blocks were not combined across the four distinct emission episodes. The time intervals between episodes and the blocks with insufficient significance were excluded. This procedure resulted in a total of 16 time bins, as shown in the bottom panel of Figure~\ref{fig:multi_panel}.

We performed spectral fitting for each time bin using the PL+BB model and, for comparison, also fitted the Band function. Detailed fit parameters for both models are provided in the supplementary material \citep{swain_2025_17782846}\footnote{\url{https://zenodo.org/records/17782846}}.

In our analysis, we find that the PL+BB model generally provided a better fit than the Band model across most time bins. However, in bin 8 of duration $\sim$16~s during Episode~3, the Band model showed a slightly better fit compared to the PL+BB model. This difference in their performance is indicated by the lower values of AIC and BIC, with differences of  $\Delta\mathrm{AIC} = 12$ and $\Delta\mathrm{BIC} = 12$. The spectra for this bin revealed the highest peak energy of $E_{\rm p} = 971^{+112}_{-108}$~keV and a BB temperature of $66.56 \pm 5.37$~keV. This suggests that, during this interval, the prompt emission is increasingly dominated by nonthermal processes, while a thermal component may still be present but is less statistically favored.

\vbdone In the Band model, the parameter $\alpha_{\rm p}$ varies from a hard spectrum value of $-0.002$ to a soft spectrum value of $-0.97$, with an average of $-0.62$. The spectrum is hardest in bin 1, which indicates optically thick emission. The harder spectra, represented by higher values of $\alpha_{\rm p}$ in the first three bins, suggest the presence of a BB component. As the episode progresses, the spectrum softens due to the expansion of the photosphere and the dominance of the nonthermal component \citep{Ryde_2010, 2011ApJ...730..141Z}. The parameter $E_{\rm p}$ tracks the light curve intensity, ranging from 211~($\pm 23$) to 970~($\pm 108$)~keV, with the highest value in bin 8 of Episode~3. The parameter $\beta_{\rm p}$ ranges from $-3.39$ to $-2.15$. The top three panels in Figure~\ref{fig:tr_flux_evolution} illustrate the evolution of flux and spectral parameters derived from the Band model across the bins.

\vbdone For the PL+BB model, the PL index $\Gamma_{\rm PL}$ lies in the range from $-1.4$ to $-1.7$, while the BB temperature ($kT$) ranges from 39.6~($\pm 2.5$) to 121~($\pm 7$)~keV. Both the peak energy and the BB temperature broadly track the intensity of the light curve and show a similar trend. We find that while the PL contributes more than half the flux in the full 10~keV--38~MeV range, the BB component dominates in the 10--1000~keV range (bottom four panels in Figure~\ref{fig:tr_flux_evolution}).

In contrast to the work of G26, which focused on Band and CPL models for their time-resolved analysis, our findings suggest that the PL+BB model is statistically preferred in most time bins. This model yields better constraints on the parameters, making it more suitable for exploring the photospheric emission and the physical properties of the jet.

\subsection{Analysis of photospheric emission} \label{sec:photospheric}

The thermal emission originates near the photospheric radius ($r_{\mathrm{ph}}$), where the optical depth drops to unity and photons decouple from baryon-associated electrons. In this section, we characterize the jet outflow at the photosphere. Such an analysis requires a segment of the light curve with significant photospheric emission and a freely expanding, nonmagnetized jet \citep{Ryde_2004, 2009ApJ...702.1211R}. At later times, the jet becomes optically thin, and the emission is typically dominated by nonthermal processes. Additionally, the curvature effect---arising from differences in photon arrival times and Lorentz boosting across a curved emitting surface---is more prominent in spectrally hard (thermal) bursts than in nonthermal ones \citep{2009ApJ...702.1211R}. As a result, high-latitude emission can significantly contribute at late times. Thus, we exclude Episodes 3 and 4 from our photospheric analysis.

\begin{figure*}[p]
    \centering
    \includegraphics[width=1\textwidth]{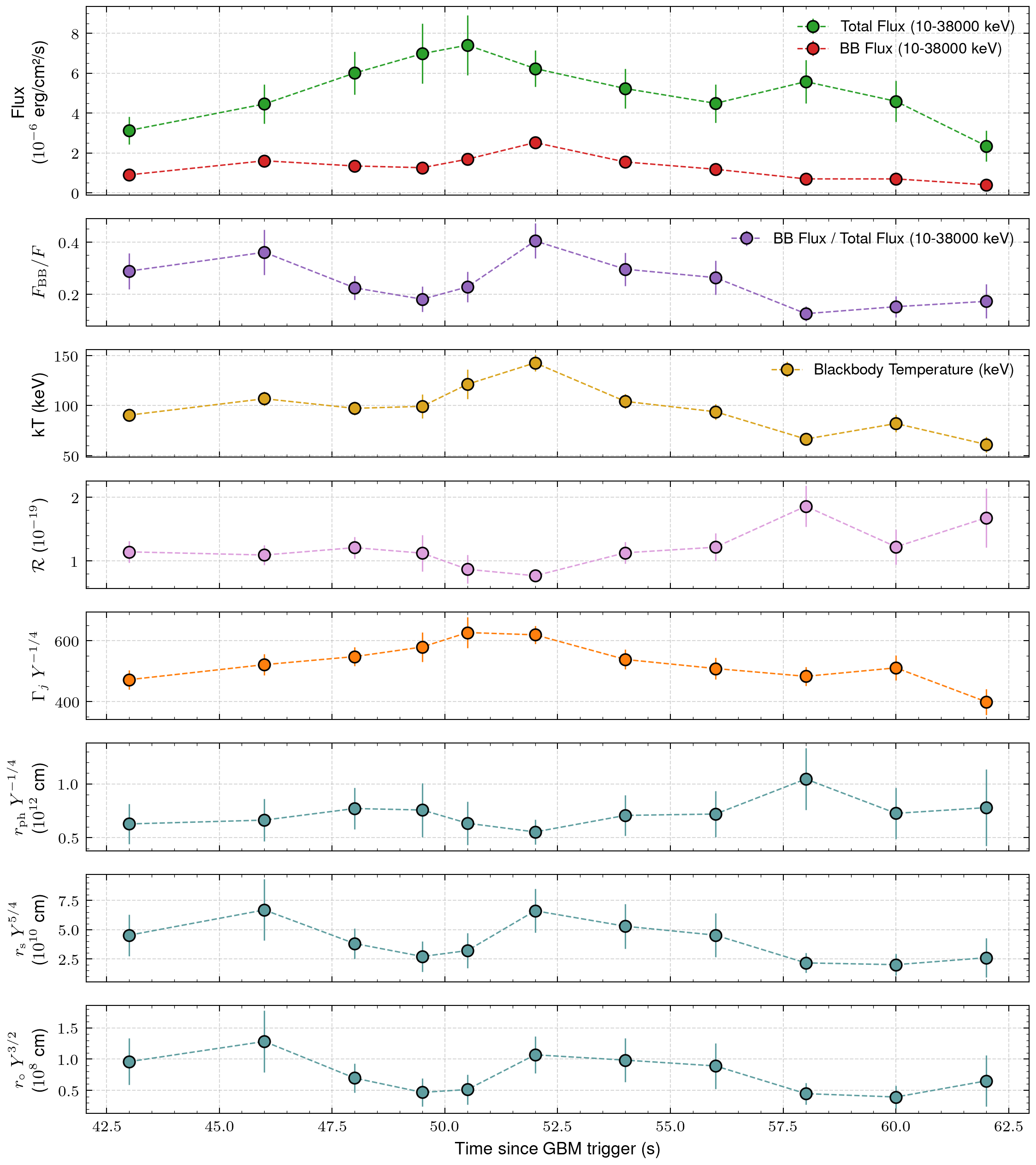}
    \caption{Time-resolved analysis of the photospheric emission during Episode~2 of \thisgrb. From top to bottom, the panels show (1) total and BB flux in the energy range of 10-38000~keV, (2) the fractional BB contribution, (3) BB temperature ($kT$), (4) effective transverse size of the emitting region ($\mathcal{R}$), (5) bulk Lorentz factor ($\Gamma_{j}$), (6) photospheric radius ($r_{\mathrm{ph}}$), (7) saturation radius ($r_{\mathrm{s}}$), and (8) base radius of the flow ($r_{\rm 0}$). The evolution of these parameters reflects the dynamics of the outflow at the photosphere.}
    \label{fig:photospheric}
\end{figure*}

The first episode of \thisgrb lasts for 6.5~s and shows significant photospheric emission; however, the signal-to-noise ratio is not high enough for a time-resolved analysis. In contrast, Episode~2 is both bright and prolonged, lasting approximately 34~s. This makes it an ideal candidate for detailed photospheric modeling using finer temporal binning.

\vbdone We follow the method described in \citet{Peer_2007}, with assumptions supported by \citet{10.1093/mnras/stt863} and \citet{Ryde_2010}. We further divide Episode~2 into 11 smaller time bins, such that the outflow can be approximated as quasi-static within each bin. The time-resolved analysis is performed using a PL+BB model, and the resulting data are available in the supplementary material at \cite{swain_2025_17782846}\footnote{\url{https://zenodo.org/records/17782846/files/Table_PL+BB-photospheric-analysis.csv}}. For the photospheric analysis, we model the outflow as adiabatic, following standard fireball dynamics under the assumption of quasi-thermal emission. In each time bin, we assume that a thermal component coexists with a nonthermal component. The evolution of the BB flux ($F_{\mathrm{BB}}$), total flux ($F$), ratio $F_{\mathrm{BB}}/F$, and the BB temperature ($kT$) are displayed in the upper panels of Figure~\ref{fig:photospheric}.

\vbdone Using the observed BB flux and temperature, we can define the ratio ($\mathcal{R}$) as follows:
\begin{equation}
\mathcal{R} = \left(\frac{F_{\mathrm{BB}}}{\sigma_{\rm SB} (kT)^4}\right)^{1/2},
\end{equation}
where $\sigma_{\rm SB}$ is the Stefan--Boltzmann constant. For \thisgrb, $\mathcal{R}$ ranges from $(0.76-1.86) \times 10^{-19}$, remaining nearly constant at $\sim 10^{-19}$ during the early rising phase of the light curve and rising towards the end of the episode (Figure~\ref{fig:photospheric}).

The photospheric radius ($r_{\rm ph}$) and the bulk Lorentz factor ($\Gamma_j$) are two important parameters that characterize the outflow at the photosphere. In the fireball model, plasma expands from the initial radius ($r_{\rm 0}$), accelerating until it reaches the saturation radius ($r_{\rm s}$). Beyond this point, the bulk Lorentz factor becomes constant, represented as $\Gamma_j = L/\dot{M}c^2$, where $L$ is the isotropic equivalent burst luminosity and $\dot{M}$ is the mass ejection rate \citep{Peer_2007}. 

The location of the photosphere ($r_{\rm ph}$) relative to $r_{\rm s}$ determines whether physical parameters like $\Gamma_j$ and $r_{\rm ph}$ can be reliably estimated. If $r_{\rm ph} < r_{\rm s}$, the flow is still accelerating with $\Gamma_j(r) \propto r$, and the relation $r_{\rm ph}/\Gamma_j = r_{\rm 0}$ holds, but the individual values of $\Gamma_j$ and $r_{\rm ph}$ remain unconstrained.

\vbdone If the BB flux is subdominant to the total gamma-ray flux ($F$), i.e., $F_{\mathrm{BB}} \ll F$, then $r_{\mathrm{ph}}$ will be less than $r_{\rm s}$ \citep{Peer_2007, 10.1093/mnras/stt863}. For \thisgrb, the ratio $F_{\mathrm{BB}}/F$ ranges from 0.12 to 0.40 (shown in Figure~\ref{fig:photospheric}), allowing reliable estimation of $\Gamma_j$ and $r_{\mathrm{ph}}$. In such a case, the bulk Lorentz factor is given by
\begin{equation}
    \Gamma_j \simeq \left(\frac{(1+z)^2 D_L Y F \sigma_{T}}{
    2 m_p c^3 \mathcal{R}
    }\right)^{1/4},
    \label{eq:Gamma}
\end{equation}
where $D_L$ is the luminosity distance, $F$ is the total observed flux, $\sigma_{T}$ is the Thomson cross section, $m_p$ is the proton mass, and $Y$ is the ratio of total fireball energy to observed gamma-ray energy, which ranges from 1 to 2. A lower value of $Y$ indicates that the efficiency of converting the kinetic energy of the outflow into the nonthermal component is very high. Thus, for energetic GRBs, $Y$ is expected to be 1. The photospheric radius ($r_{\rm ph}$) can be estimated as
\begin{equation}
    r_{\rm ph} \simeq  \frac{L~\sigma_{T}}{8\pi \Gamma_j^{3} m_p c^3},
\end{equation}
where $L = 4\pi D_L^{2}YF$ is the isotropic equivalent burst luminosity.

\vbdone As seen in Figure~\ref{fig:photospheric}, for \thisgrb\ the bulk Lorentz factor ranges from 400 to 626 and tracks the total flux intensity. On the other hand, the photospheric radius is almost constant around $(0.5-1.0) \times 10^{12}~\mathrm{cm}$.

\vbdone The radius at the base of the outflow ($r_{\rm 0}$), which is independent of the cases ($r_{\rm ph} < r_{\rm s}$ and $r_{\rm ph} > r_{\rm s}$), is given by
\begin{equation}
    r_{\rm 0} \simeq \frac{D_L \mathcal{R}}{(1+z)^2} \left(\frac{F_{\rm BB}}{Y F}\right)^{3/2}. 
\end{equation}
Finally, the saturation radius is defined as $r_{\rm s} = \Gamma_j\times r_{\rm 0}$, which depends on $\Gamma_j$. Thus, it can be defined for the case where $r_{\rm ph} > r_{\rm s}$.

\vbdone For \thisgrb, the base radius of the outflow, $r_{\rm 0}$, ranges from $(0.4-1.3) \times 10^{8}~\mathrm{cm}$, and the saturation radius ranges from $(2.0 - 6.7) \times 10^{10}~\mathrm{cm}$ (indicating $r_{\rm s} < r_{\rm ph}$). We find that both $r_{\rm 0}$ and $r_{\rm s}$ strongly depend on the thermal contribution to the total flux. 

\section{Prompt emission: interpretation} \label{sec:prompt_interpretation}

\subsection{Photospheric emission} \label{sec:photospheric-int}

\begin{figure*}[th]
    \centering
    \includegraphics[width=1\textwidth]{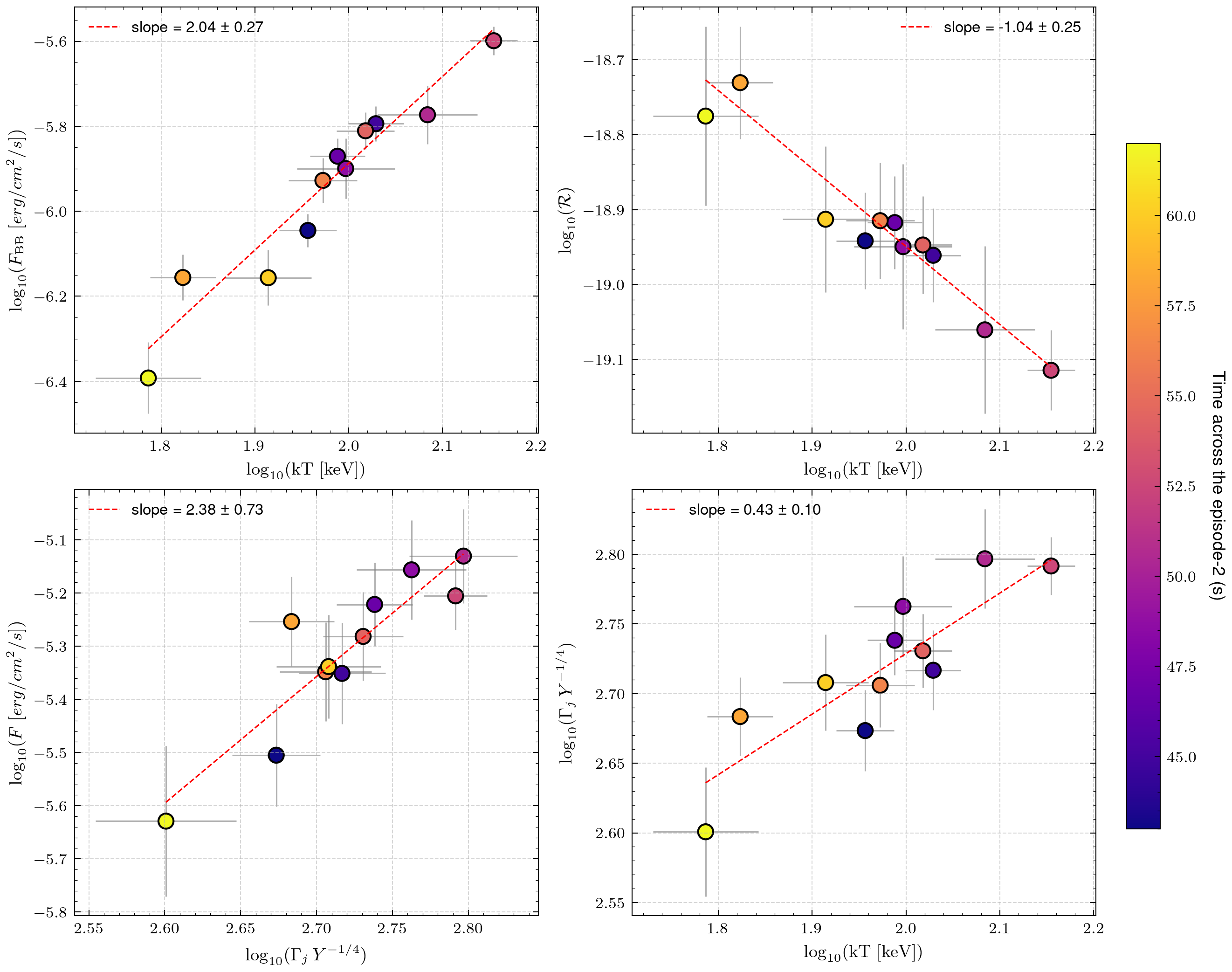}
    \caption{Correlations between photospheric parameters during Episode 2 of \thisgrb. The upper panels display the evolution of BB flux ($F_{\rm BB}$) and the ratio $\mathcal{R}$ as functions of BB temperature ($kT$). The lower-left panel shows the correlation between the total flux ($F$) and the bulk Lorentz factor ($\Gamma_j$), while the lower-right panel presents the relation between the Lorentz factor and BB temperature. The color bar indicates the temporal evolution across the episode.}
    \label{fig:slop_fit_plot}
\end{figure*}

The time-resolved photospheric parameters obtained for Episode~2 provide crucial insights into the physical state and evolution of the relativistic outflow. In the lower-left panel of Figure \ref{fig:slop_fit_plot}, we illustrate the relationship between the total flux and the bulk Lorentz factor, which follows the dependence $F \propto \Gamma_j^{(2.38 \pm 0.73)}$. This relationship indicates that the total radiated flux is sensitive to changes in the Lorentz factor, consistent with expectations from emission due to internal shocks.

Similarly, the lower-right panel of Figure \ref{fig:slop_fit_plot} shows the correlation between the BB temperature and the Lorentz factor, following $\Gamma_j \propto T^{(0.43 \pm 0.10)}$. This suggests that the thermal and dynamical properties of the outflow evolve coherently during the prompt emission phase. 

Conversely, the photospheric radius remains remarkably stable compared to the variations in temperature and flux. A similar characteristic was observed in the case of the thermally bright GRB~090902B \citep{Ryde_2010}, supporting the interpretation that such bursts share a common photospheric emission mechanism.

We find that $\mathcal{R}$ is negatively correlated with the BB temperature ($kT$) following $\mathcal{R} \propto kT^{-1.04 \pm 0.25}$ for \thisgrb, as shown in the upper right panel of Figure~\ref{fig:slop_fit_plot}. As a consequence, despite the effective emitting area increasing at late times, the overall contribution of the BB flux declines due to the decreasing photospheric temperature. This trend is consistent with an expanding and cooling photosphere, which gradually becomes less significant as nonthermal emission processes become more prominent.

Overall, the photospheric properties of Episode~2 of \thisgrb\ are consistent with a relativistic outflow that is initially weakly magnetized and dominated by quasi-thermal emission at early times. The gradual decline in thermal dominance, together with the stability of the photospheric radius and the correlated evolution of temperature and Lorentz factor, points to a transition from photospheric to nonthermal emission as the jet evolves and the photosphere cools. This behavior supports a scenario in which the jet is launched deep within a stellar progenitor and propagates adiabatically before becoming optically thin, giving rise to the observed interplay between thermal and nonthermal components in the prompt emission phase of the long GRBs.

\subsection{Origin of the nonthermal component}
\vbdone In this section, we discuss three possible mechanisms responsible for producing nonthermal emission in this GRB and their connection to the observed photospheric component.

One possibility is internal shocks, where collisions between relativistic shells dissipate kinetic energy and produce nonthermal emission, typically via synchrotron radiation. Since both the thermal and nonthermal components are powered by the central engine, they are expected to be temporally correlated \citep{Ryde_2010}, consistent with the behavior observed during the early episodes of \thisgrb\ (Figure~\ref{fig:tr_flux_evolution}).

Alternatively, the correlation between the PL and BB components can be explained by inverse Compton processes, such as synchrotron self-Compton emission or the Compton upscattering of thermal photons \citep{2011ApJ...730..141Z}. In this context, the low-energy power law component might originate from a different emission mechanism, such as synchrotron radiation. The positive slope of the PL component in $\nu F_{\nu}$  space suggests that a significant fraction of the emitted energy could be in the GeV range or higher. While the \emph{Fermi} Large Area Telescope (LAT) has detected such emissions in thermally bright GRB (e.g., GRB~090902B; \citealt{Ryde_2010}), LAT observations for \thisgrb\ are not available due to its large boresight angle.

\vbdone A third possibility is that the nonthermal PL component may arise from magnetic reconnection, where energy is transported via Poynting flux within the baryonic outflow. GRBs of this type, such as GRB~080916C, are expected to possess highly magnetized jets with purely nonthermal emission that is well described by the Band function \citep{2011ApJ...732...49P}. For \thisgrb, the presence of a BB component in the observed spectrum indicates that the jet is either moderately or weakly magnetized \citep{Veres_2012}, making this interpretation less likely.

\subsection{Empirical correlation of global parameters}\label{sec:empirical_correlation}
\vbdone GRBs exhibit several empirical correlations among prompt emission observables, which are often used to characterize the common behavior of the jet outflows and constrain the physical properties of their progenitors. Here, we compare the prompt emission properties of \thisgrb\ with those of (energetic) LGRBs using rest-frame quantities derived from the time-integrated Band function parameters.

To enable a consistent comparison, we compute rest-frame quantities in the standard 1--$10^{4}$~keV energy band using a $k$-correction. Detailed derivations are provided in Appendix~\ref{kcorrection}. For \thisgrb, we obtain a $k$-correction factor of $\sim0.97$ for the observed 10--38,000~keV band. Using a fluence of $S_\gamma = (1.9 \pm 0.2)\times10^{-4}$~erg~cm$^{-2}$ and a luminosity distance of $D_L = 5.34\times10^{28}$~cm, we derive an isotropic equivalent energy of $E_{\rm iso} = (2.2 \pm 0.4)\times10^{54}$~erg. Similarly, from the observed peak energy $E_{\rm p} = 560 \pm 55$~keV, we calculate the rest-frame peak energy $E_{\rm p,z} = E_{\rm p} \times (1+z) = 1758 \pm 172~\mathrm{keV}$.

We note that although \thisgrb\ follows both the Amati and Yonetoku relations, its isotropic equivalent energy is significantly higher than that of typical LGRBs, ranking it among the most energetic events in this class. GRBs dominated by photospheric emission are expected to naturally follow the Amati relation \citep{Lazzati_2013}, which aligns with our observations for \thisgrb, where a prominent thermal component is detected during the prompt phase. These empirical correlations are also discussed in G26. In Appendix~\ref{appendix_amati}, we provide a comparison of samples in the context of energetic long GRBs, along with the relevant plots.

We also estimate the initial bulk Lorentz factor ($\Gamma_{j,0}$) using the empirical Liang correlation \citep{Liang_2010}, which relates the isotropic equivalent gamma-ray energy to the initial Lorentz factor of the outflow:
\begin{equation}
\Gamma_{j,0} \simeq 182 \left( \frac{E_{\mathrm{iso}}}{10^{52}\ \mathrm{erg}} \right)^{0.25 \pm 0.03}.
\label{gamma_cal}
\end{equation}
Using the measured value of $E_{\mathrm{iso}}$, we obtain $\Gamma_{j,0} \sim 700$ for \thisgrb. This estimate is consistent with the bulk Lorentz factors inferred independently from the time-resolved photospheric analysis (Section~\ref{sec:photospheric}), supporting a highly relativistic outflow.

\citet{2001ApJ...562L..55F} found that in pre-Swift GRBs, the jet opening angle $\theta_0$ seems to be anticorrelated with $E_{\mathrm{iso}}$ through $E_{\rm iso} \propto \theta^{-2}_0$, so that the jet-corrected gamma-ray energy $E_{\gamma} \approx (\theta^2_0 / 2)E_{\mathrm{iso}}$ is roughly constant, $\approx 10^{51}$~erg for LGRBs. \added{Subsequent studies have shown that this characteristic energy can extend up to $\sim10^{52}$~erg, particularly for energetic GRBs \citep{Cenko_2010}. The implication is that wider jets tend to have a lower energy concentration, while narrow jets have a higher energy concentration. Using this relation, we estimate a jet opening angle of $\theta_0 \sim (1.7^\circ - 5.5^\circ)$, suggesting a relatively collimated outflow.}

\section{Afterglow emission: analysis}\label{sec:afterglow}
The afterglow of \thisgrb\ has been detected from X-ray to radio wavelengths. In the optical band, we have well-sampled data up to $\sim2.3$~days after the burst, with the most extensive coverage obtained in the SDSS $r$ filter. The $r$-band light curve is described by a single PL decay with a temporal index of $\alpha = 1.82 \pm 0.01$. The optical light curve displays a bright and rapidly decaying afterglow, associated with an energetic GRB, making it an intriguing event for study. However, our data have limited multiband coverage, which makes it difficult to constrain the spectral properties of this GRB.

In the radio frequencies, the afterglow was first detected by ATCA in the 16.7 and 21.2~GHz bands approximately 3.6~days after the burst. We later conducted additional radio observations using GMRT and VLA, which provided critical late-time coverage. In particular, our VLA observations at the 10~GHz band detected emissions at 10.4 and 87.4 days following the burst. The late-time radio emission exhibits a plateau-like behavior, contrasting with the rapid fading of the optical afterglow.

In the X-ray band, only two data points are available from \emph{Swift}-XRT, taken approximately 1 and 9.8~days after the burst. The source was very faint, with no reliable centroid determination, and the final data point has marginal significance, limiting its utility for afterglow modeling.

\begin{figure}[ht!]
    \centering
    \includegraphics[width=\columnwidth]{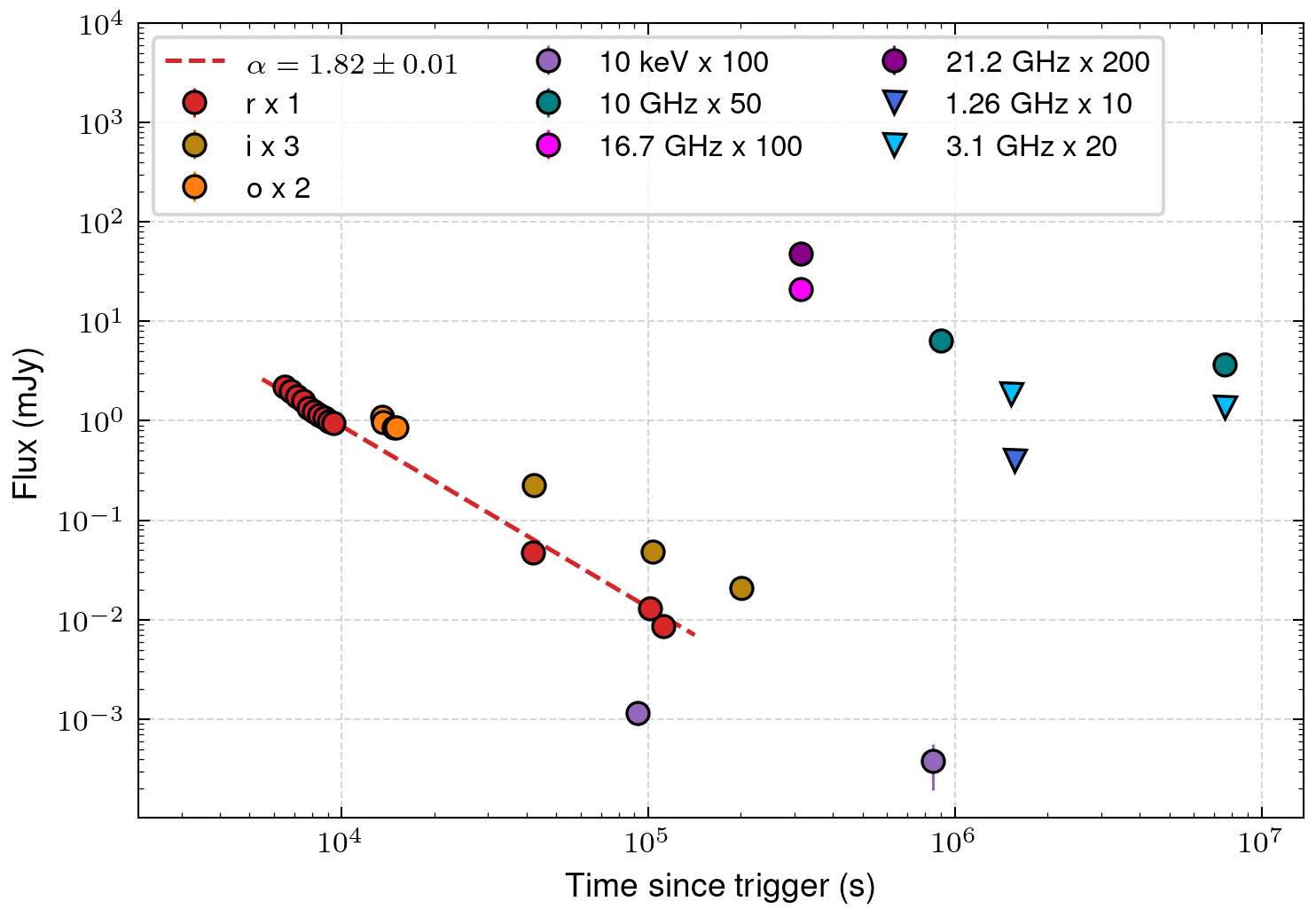}
    \caption{Multiwavelength afterglow light curves of \thisgrb. Optical, X-ray, and radio data are shown; photometry has not been corrected for host-galaxy extinction. The $r$-band optical light curve exhibits a rapidly fading afterglow, well described by a single PL decay.}
    \label{fig:LC}
\end{figure}

All the afterglow data used in this work are available in the supplementary material at \cite{swain_2025_17782846}\footnote{\url{https://zenodo.org/records/17782846/files/Table_afterglow-obs.csv}} and are plotted in Figure~\ref{fig:LC}.

Overall, the afterglow of \thisgrb\ is characterized by a steep optical decay at early times and a long-lived radio component at late times, providing valuable constraints for broadband afterglow modeling.

\section{Afterglow modeling}\label{sec:afterglow_model}
In this section, we model the afterglow emission to constrain the macrophysical parameters of the outflow, including the isotropic equivalent kinetic energy ($E_{\rm K,iso}$), jet opening angle, and observer viewing angle, as well as microphysical parameters such as the electron PL index ($p$), the fraction of energy in electrons ($\epsilon_e$), the fraction in magnetic fields ($\epsilon_b$), and the fraction of particles accelerated into a nonthermal distribution ($\chi$). The temporal evolution of the afterglow depends on the ordering of the observing frequency $\nu$ relative to the characteristic synchrotron frequency $\nu_m$ and the cooling frequency $\nu_c$ \citep{Sari_1997, 2000ApJ...536..195C, Granot_2002}.

For the uniform density medium, the observed optical decay index of $\alpha = 1.82$ corresponds to $p \simeq 3.4$ in the slow-cooling regime ($\nu_m < \nu < \nu_c$), which is significantly higher than the typical range ($2 \lesssim p \lesssim 3$) expected from shock-acceleration theory and commonly inferred from GRB afterglow observations \citep{Wang_2015, 2022Galax..10...66M}. Likewise, the nearly constant late-time radio flux is inconsistent with the expectations for a uniform density medium. We therefore rule out the uniform density CBM within the standard forward-shock framework for this event.

G26 performed broadband afterglow modeling, assuming both top-hat and structured jets interacting with either a uniform density or a wind-type CBM, and found that a top-hat jet provided a better description of the data. However, in both environments, their best-fit models predict relatively shallow temporal decay indices ($\alpha \sim 1.21$ for a uniform medium and $\alpha \sim 1.34$ for a wind-like medium) in the slow-cooling regime ($\nu_m < \nu < \nu_c$), which underestimate the steep optical decay ($\alpha = 1.82$) observed for \thisgrb. In the next section, we proceed with the afterglow modeling by assuming a wind-type CBM.

\subsection{Wind-type medium}\label{sec:afterglow_wind}
\begin{figure}[ht!]
    \centering
    \includegraphics[width=\columnwidth]{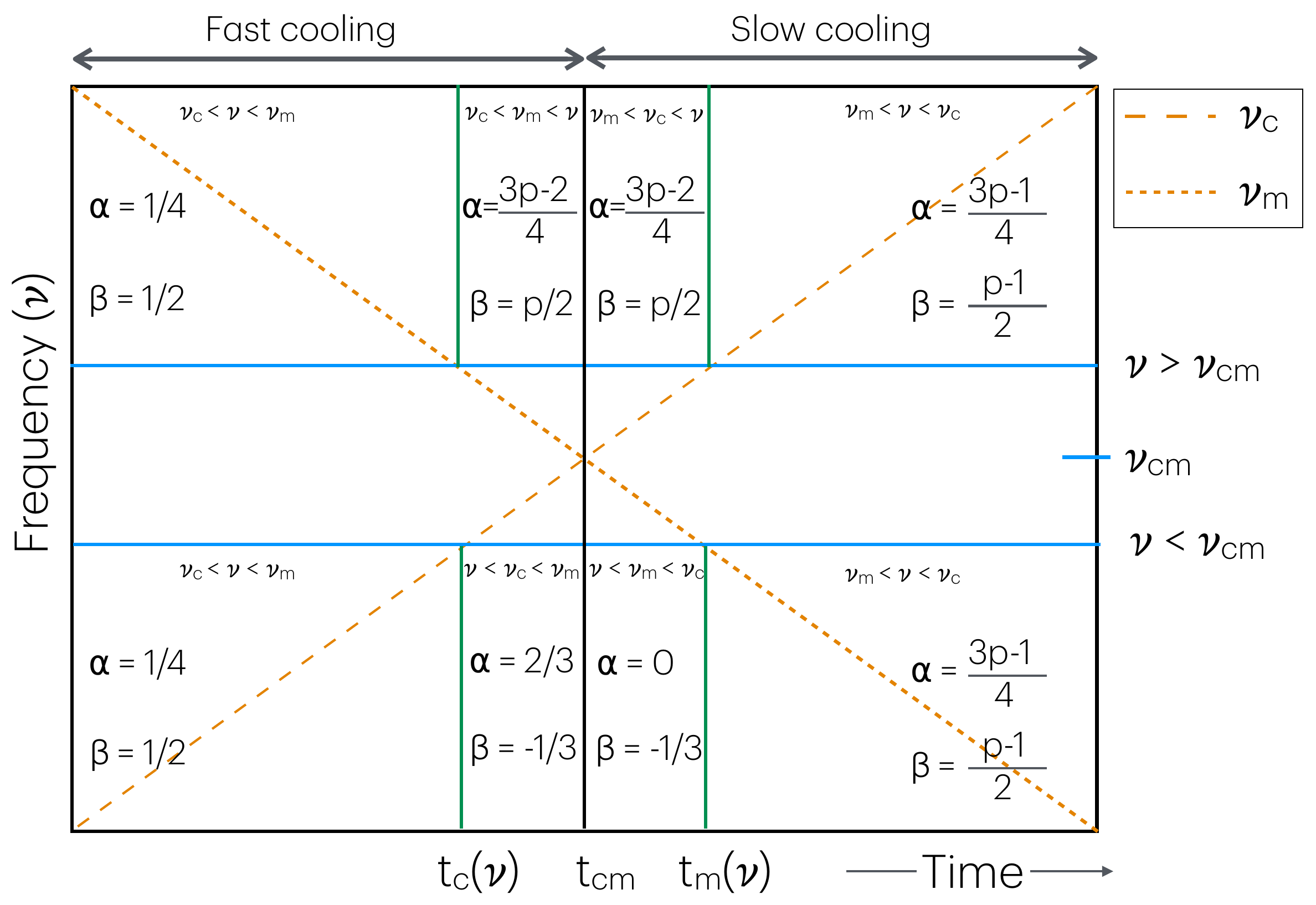}
    \caption{Schematic showing closure relations for $\alpha$ and $\beta$ following $F \propto t^{-\alpha}\nu^{-\beta}$ under different regimes for a wind-type medium. The frequency--time space is divided into regimes based on the evolution of the synchrotron characteristic and cooling frequencies as a function of time. The relevant closure relation is identified based on the observation frequency and time. Please note that the trends are indicative, and the figure is not drawn to scale. Based on calculations and discussions in \citet{2000ApJ...536..195C}.}
    \label{fig:wind_closure_rel}
\end{figure}

\citet{2000ApJ...536..195C} extensively discusses the afterglow of a GRB exploding in a wind-type medium. In this section, we examine the relevant closure relations for such a medium, summarized schematically in Figure~\ref{fig:wind_closure_rel}. The dashed diagonal line denotes the cooling frequency increasing with time ($\nu_c \propto t^{1/2}$), while the dotted diagonal shows the characteristic synchrotron frequency decreasing with time ($\nu_m \propto t^{-3/2}$). The epoch at which these frequencies are equal ($\nu_c = \nu_m = \nu_{cm}$) marks the transition from the fast- to slow-cooling regime and typically occurs at very early times \citep{1998ApJ...497L..17S}. Since the optical afterglow of \thisgrb\ exhibits a single PL decay without any detectable temporal break, we restrict our analysis to the slow-cooling regime ($t > t_{cm}$).

In the X-ray band, the afterglow emission generally satisfies $\nu > \nu_{cm}$. In this case, the temporal decay is expected to follow $\alpha = (3p-2)/4$ for $\nu_m < \nu_c < \nu$, with a mild steepening to $\alpha = (3p-1)/4$ ($\nu_m < \nu < \nu_c$) once $\nu_c$ crosses the observing band.

In the optical band, two scenarios are possible: either $\nu > \nu_{cm}$, analogous to the X-ray regime, or $\nu < \nu_{cm}$. In the latter case, the early-time light curve is expected to exhibit a plateau phase ($\nu < \nu_m < \nu_c$), followed by a steeper decay with $\alpha = (3p-1)/4$ ($\nu_m < \nu < \nu_c$) once $\nu_m$ drops below the observing frequency. Notably, the late-time decay index is identical in both scenarios. This degeneracy can be resolved by observing the afterglow over a broad wavelength range, as the passage of spectral break frequencies ($\nu_m$ or $\nu_c$) through different bands creates distinct temporal signatures that help constrain the frequency ordering.

The optical afterglow of \thisgrb\ follows a single PL decay out to $\sim30.8$~hr, indicating that the observations likely probe the final stages of the decay phase ($\nu_m < \nu < \nu_c$). Under this assumption, we infer an electron PL index of $p = 2.76 \pm 0.01$, consistent with expectations from the synchrotron forward-shock model.

At radio frequencies, synchrotron self-absorption plays a crucial role in shaping the light curve. In a wind-type medium, the self-absorption frequency $\nu_a$ decreases with time and eventually crosses the characteristic frequency $\nu_m$. However, this transition typically occurs at very late times \citep{2022ApJ...927...84Z}. Consequently, during the epochs probed by our radio observations, the afterglow is expected to satisfy $\nu < \nu_{cm}$. In this case, the radio light curve will exhibit a nearly constant radio flux, followed by a late-time steeper decay, a hallmark of a wind-type environment. This behavior is indeed observed in our data (Figure~\ref{fig:LC}).

\subsection{Semi-analytical solutions}
\vbdone Based on the various considerations in \S\ref{sec:afterglow_wind}, we now fit our afterglow data to calculate the physical parameters of the outflow. The additional parameter $A_\star$ characterizes the density normalization of the wind-type CBM, following the definition from \cite{2000ApJ...536..195C}. We assume an adiabatic top-hat jet with a high initial Lorentz factor, observed within the jet opening angle. In our modeling, we fix $\chi = 1$, corresponding to efficient particle acceleration. As no jet break is detected in the available data, the jet opening angle remains weakly constrained.

For a given set of physical parameters, the characteristic synchrotron frequencies $\nu_m$, $\nu_c$, and $\nu_a$ are computed, allowing the model flux to be evaluated as a function of time and observing frequency, following \citet{Granot_2002}. We infer the model parameters using nested sampling via the \sw{MultiNest} algorithm, implemented through the \sw{PyMultiNest} package \citep{2014A&A...564A.125B}, with 2000 live points. Uniform priors for the model parameters are listed in Table~\ref{tab:mcmc_post}. The posterior distributions are obtained by maximizing the likelihood of the model given the observed data, and the resulting parameter estimates and credible intervals are reported in Table~\ref{tab:mcmc_post} (Posterior-1). 

In our initial run, we found significant degeneracies among the model parameters. The resulting posterior distributions span a wide range of values: $E_{K,\mathrm{iso}} \in (10^{54} - 10^{57})~\mathrm{erg}$, $\epsilon_e \in 0.03 - 0.23$, $\epsilon_b \in 10^{-6} - 10^{-4}$, and $A_\star \in 0.01 - 0.77$, all within the 95\% confidence interval. However, the posterior for the electron energy index, $p$, is well constrained, yielding $p = 2.76 \pm 0.01$, and is consistent with the analytic estimate derived in \S\ref{sec:afterglow_wind}.

To reduce parameter degeneracies, we adopt a typical value of $\epsilon_e \sim 0.1$ \citep{2002ApJ...571..779P, Cenko_2010}. For the parameter $\epsilon_b$, \citet{Zhang_2007} place an upper limit of $\epsilon_b \lesssim 10^{-4}$ for X-ray afterglows in the regime $\nu_m < \nu_{\rm X} < \nu_c$, which is expected for a wind-like CBM. We therefore explore two representative values of $\epsilon_b$, $10^{-4}$ and $10^{-5}$, in our subsequent analysis.

With $\epsilon_e$ and $\epsilon_b$ fixed, the posterior samples yield $E_{K,\mathrm{iso}} = (2.7 \pm 0.5) \times 10^{54}\ \mathrm{erg}$ for $\epsilon_b = 10^{-4}$ and $E_{K,\mathrm{iso}} = (2.7 \pm 0.5) \times 10^{55}\ \mathrm{erg}$ for $\epsilon_b = 10^{-5}$. In both cases, we obtain $A_\star = 0.15 \pm 0.02$ and $p = 2.76 \pm 0.01$. The results are summarized in the Posterior-1 and Posterior-2 columns of Table~\ref{tab:mcmc_post}.

\begin{figure}[ht]
    \centering
    \includegraphics[width=\columnwidth]{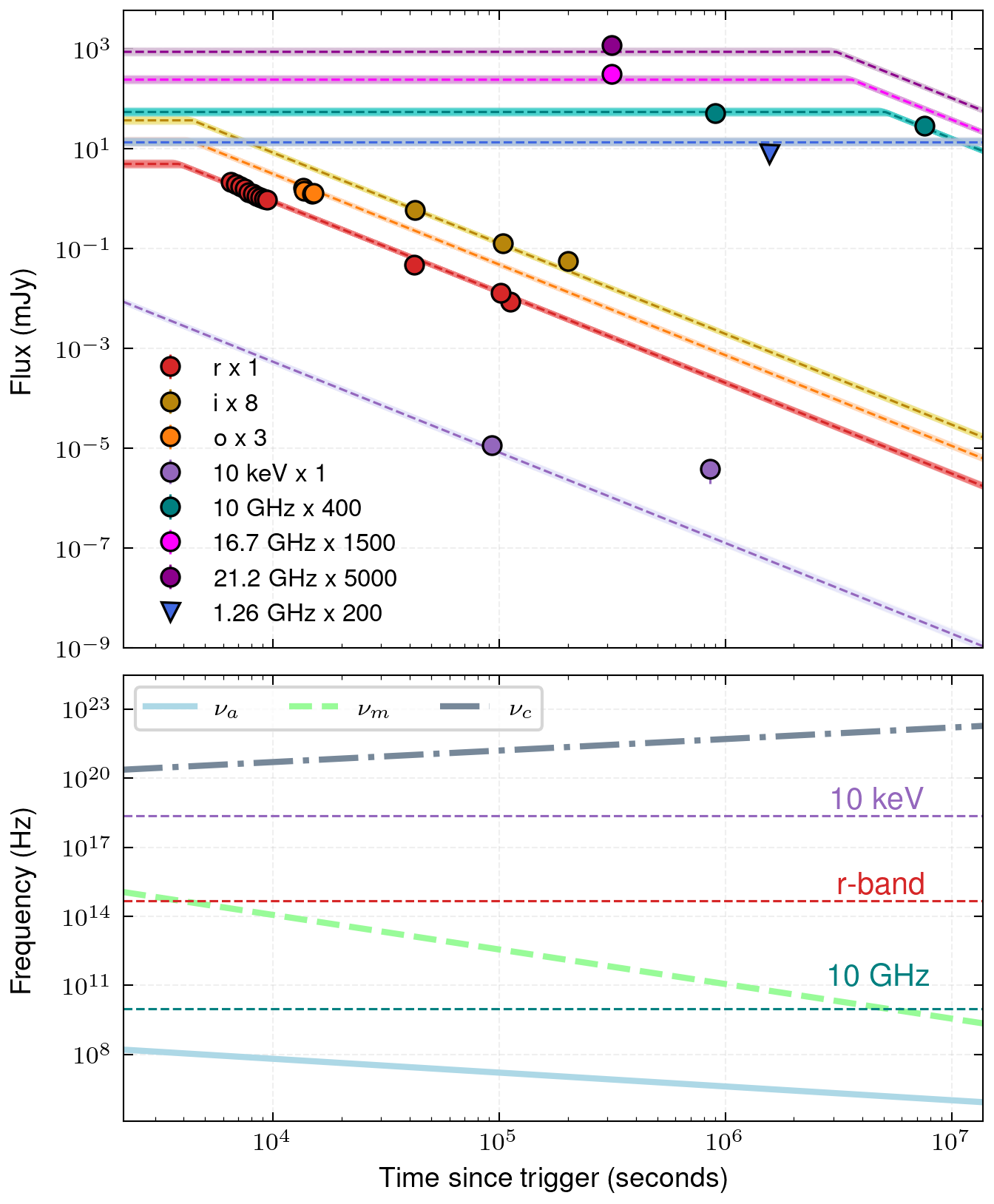}
    \caption{Afterglow modeling of \thisgrb\ with an afterglow model assuming a relativistic top-hat jet propagating into a wind-type medium with the observer's line of sight located within the jet opening angle. The upper panel shows the multiband afterglow light curves plotted with dotted lines representing the best-fit model for each band. The shaded regions indicate the $95\%$ credible intervals from the fit. Bottom panel shows the synchrotron break frequencies $\nu_a$, $\nu_m$ and $\nu_c$ as a function of time.}
    \label{fig:modeled_LC}
\end{figure}

The inferred light curve for the best-fit model corresponding to $\epsilon_e = 0.1$ and $\epsilon_b = 10^{-4}$, overlaid on the observed data, is shown in the upper panel of Figure~\ref{fig:modeled_LC}. The dashed line represents the model light curve based on the median parameter values, while the shaded region indicates the 95\% credible interval. The temporal evolution of the characteristic frequencies, $\nu_a$, $\nu_m$, and $\nu_c$, computed from the median parameters, is shown in the lower panel. As discussed in \S\ref{sec:afterglow_wind}, the characteristic frequency $\nu_m$ crosses the optical band at very early times, before our first observation, and passes through the 10~GHz band at $\sim48$~days after the explosion, resulting in a slight decrease in the radio flux. The cooling frequency $\nu_c$ remains above all observational bands throughout the period of interest. 

For a wind-like medium, the inferred kinetic energy depends sensitively on the assumed microphysical parameters, particularly $\epsilon_e$, $\epsilon_b$, and $A_\star$, leading to substantial uncertainties in the radiative efficiency. For \thisgrb, adopting $\epsilon_b \lesssim 10^{-4}$ and $E_{K,\mathrm{iso}} \gtrsim 2.7 \times 10^{54}$~erg yields an upper limit on the radiative efficiency of $\sim45\%$. Lower values of $\epsilon_b$ imply larger kinetic energies and correspondingly lower efficiencies; for example, adopting $\epsilon_b = 10^{-5}$ yields $E_{K,\mathrm{iso}} \sim 2.7 \times 10^{55}$~erg and a radiative efficiency of $\sim7.5\%$. Overall, our modeling suggests a moderately broad efficiency range that is consistent with values typically inferred for energetic GRBs.

The afterglow modeling by G26 inferred a lower magnetic field fraction ($\epsilon_b \sim 10^{-6}$) and a higher kinetic energy ($E_{K,\mathrm{iso}} \sim 4 \times 10^{55}$~erg), resulting in a lower radiative efficiency of $\sim4\%$. The differences between these estimates highlight the sensitivity of radiative efficiency to assumptions about the circumburst environment and the adopted microphysical parameters.

Overall, the inferred parameters and the modeled light curves provide a consistent description of the steeply decaying optical afterglow and the late-time radio plateau of \thisgrb, supporting a relativistic jet expanding into a wind-type CBM. Our best-fit model yields $p = 2.76 \pm 0.01$, $\epsilon_e = 0.1$, and $A_\star = 0.15 \pm 0.02$, with an upper limit of $\epsilon_b \lesssim 10^{-4}$ and a lower limit on the isotropic equivalent kinetic energy of $E_{K,\mathrm{iso}} \gtrsim 2.7 \times 10^{54}$~erg.

\begin{deluxetable*}{rcccccc}
    \tablewidth{0pt}
    \tablecaption{Priors and posteriors of the model parameters. The best-fit values are reported with their 95\% confidence intervals.  \label{tab:mcmc_post}}
    \tablehead{
    \colhead{Parameter} & \colhead{Unit} & \colhead{Prior Type} & \colhead{Parameter Bound} & \colhead{Posterior-1} & \colhead{Posterior-2} & \colhead{Posterior-3}\\
    \colhead{} & \colhead{} & \colhead{} & \colhead{} & \colhead{} & \multicolumn{2}{c}{(Fixed $\epsilon_e$ and $\epsilon_b$)}
    }
    \startdata
    $\log_{10}(E_{K,\mathrm{iso}})$ & erg & Uniform & [53, 58] & $55.79^{+1.96}_{-1.67}$ & $54.43^{+0.07}_{-0.07}$ & $55.43^{+0.07}_{-0.07}$  \\
    $\log_{10}(A_{\mathrm{*}})$ & --- & Uniform & [-2, 0] & $-1.01^{+0.90}_{-0.78}$ & $-0.83^{+0.06}_{-0.06}$ & $-0.83^{+0.07}_{-0.06}$ \\
    $\log_{10}(\epsilon_{\mathrm{e}})$ & --- & Uniform & [-2.0, -0.5] & $-1.09^{+0.45}_{-0.39}$ & -1 & -1 \\
    $\log_{10}(\epsilon_{\mathrm{b}})$ & --- & Uniform & [-6, -1] & $-5.10^{+1.03}_{-0.85}$ & -4 & -5 \\
    $p$ & --- & Uniform & [2, 3] & $2.76 \pm 0.01$ & $2.76 \pm 0.01$ & $2.76 \pm 0.01$ \\
    \enddata
\end{deluxetable*}

\section{Discussion}\label{sec:discussion}
The multiwavelength observations of \thisgrb\ reveal an energetic prompt emission and a rapidly evolving afterglow with distinct optical and radio behavior. In this section, we discuss the physical implications of our results and place \thisgrb\ in the broader context of correlations between prompt and afterglow properties.

\subsection{Fast fading afterglow in a wind-type medium} \label{sec:fast_fading}

As discussed in \S\ref{sec:afterglow_wind}, the optical afterglow of \thisgrb\ exhibits a steep decay with $\alpha \simeq 1.82$ in the $\nu_m < \nu_{\rm opt} < \nu_c$ regime, significantly steeper than that observed in a typical LGRB, as shown in Figure~\ref{fig:parameter_correlation}. In both wind-type and uniform density environments, the transition from fast to slow cooling occurs at very early times. For typical parameters, the cooling frequency lies above the optical band during most of the afterglow evolution, implying that the observed optical emission is powered by electrons that do not cool efficiently. Consequently, the optical light curve of most GRBs, including \thisgrb\, predominantly traces the adiabatic slow-cooling segment $\nu_m < \nu_{\rm opt} < \nu_c$.

The temporal decay of the afterglow in this regime depends on the surrounding medium. For a uniform density medium, the expected decay index is $\alpha = (3p-3)/4$, whereas a wind-type medium predicts a steeper decay, $\alpha = (3p-1)/4$. For a given electron PL index $p$, the afterglow in a wind environment therefore decays faster by $\Delta\alpha = 0.5$ compared to the uniform density case.

\vbdone This rapid decay can be understood by examining the peak synchrotron flux $F_{\nu, \max}$, which represents the maximum emissivity near the frequency $\nu_m$ in the slow-cooling regime \citep[using the definition from][]{1998ApJ...497L..17S}. In a wind-type medium, this peak flux decreases with time (scaling roughly as $t^{-1/2}$), while in a uniform density medium, it remains constant. The physical interpretation is that, in a uniform medium, the jet interacts with a constant-density environment and continuously sweeps up new material at a steady rate. In contrast, in a wind-type environment, the density decreases as $r^{-2}$, so the jet encounters progressively less material as it expands, leading to less radiation dissipation and creating a steeper decay in the afterglow flux.

\subsection{Collimated outflow}
\vbdone While the jet remains relativistic, the observed light curve can be adequately modeled using a spherical expansion framework. However, once the jet decelerates sufficiently, edge effects become prominent due to the lateral expansion of the jet \citep{1999ApJ...525..737R, Sari_1999}. This transition results in a significantly steeper light-curve evolution during the edge expansion phase.
The edge effect becomes observable when the Lorentz factor of the jet becomes comparable to the inverse of the jet's angular half-opening angle ($\Gamma_j \sim 1/\theta_0$). Following \citet{Chevalier_2000}, the characteristic time at which edge effects become visible is:
\begin{equation}
t_{\mathrm{edge}} = \left(\frac{1+z}{2}\right) \left(\frac{\theta_0}{0.2}\right)^{4} E_{52} A_{\star}^{-1}~\mathrm{days},
\end{equation}
where $\theta_0$ is the jet half-opening angle in radians, $E_{52}$ is the isotropic equivalent kinetic energy ($E_{K,\mathrm{iso}}$) in units of $10^{52}~\mathrm{erg}$, and $A_{\star}$ is the wind parameter.

The analysis of the \thisgrb\ light curve indicates that, up to the final radio detection at $\sim 87$~days, no significant steepening in the light curve was observed. This sets a lower limit on the jet opening angle of $\theta_0 > 4^\circ.8$ for an isotropic kinetic energy of $E_{K,\mathrm{iso}} \sim 2.7 \times 10^{54}\mathrm{erg}$. This limit is consistent with our independently derived estimate of $\theta_0 \sim (1^\circ.7 - 5^\circ.5)$, obtained in \S\ref{sec:empirical_correlation}.

\vbdone \citet{Cenko_2010} found that the highly energetic GRBs can release a true energy of $E_{\gamma} \gtrsim 10^{52}\ \mathrm{erg}$. For \thisgrb, considering the collimated jet adopting $\theta_0 \sim 5^\circ$, $E_{K,\mathrm{iso}} \gtrsim 2.7 \times 10^{54}\ \mathrm{erg}$, and $E_{\mathrm{iso}} \sim 2.2 \times 10^{54}\ \mathrm{erg}$, we estimate a total energy release of $E_{\gamma} \gtrsim 1.9 \times 10^{52}\ \mathrm{erg}$. These results place \thisgrb\ among the class of hyper-energetic GRBs.

\subsection{Implications of Wind Parameters of a GRB Progenitor}
\vbdone Considering the observed wind interaction in \thisgrb, the most plausible scenario is that the progenitor of \thisgrb\ was a W-R star. Galactic W-R stars typically exhibit wind velocities of $\sim 1000~\mathrm{km~s^{-1}}$ and mass-loss rates of $\sim 10^{-5}~\mathrm{M_\odot~yr^{-1}}$ \citep{Woosley_2006}. However, our derived wind parameter $A_\star \sim 0.15$ for \thisgrb\ implies a lower mass loss rate of $(1.5 \pm 0.2) \times10^{-6}~\mathrm{M_\odot~yr^{-1}}$. Mass-loss in the W-R phase plays a critical role in determining the final fate of the star, as excessive loss can strip the core, favoring neutron star formation and preventing the formation of the massive, rapidly rotating core needed for GRB production. Moderate mass-loss rates, on the other hand, preserve sufficient core mass and angular momentum, aligning with the collapsar model \citep{Woosley_2006}. Metallicity further influences these outcomes, as low-metallicity W-R stars have reduced mass-loss rates and are more likely to retain the rapid rotation necessary to power a jet \citep{2001ApJ...550..410M, Woosley_2006}. Thus, the mass-loss rate inferred for \thisgrb\ is consistent with this theoretical framework, supporting a low-metallicity W-R progenitor that underwent moderate mass loss while retaining sufficient core mass and angular momentum to produce a successful GRB.

\subsection{Correlation between Prompt and Afterglow Parameters} \label{sec:afterglow_corr}

\begin{figure*}[th]
    \centering
    \includegraphics[width=1\textwidth]{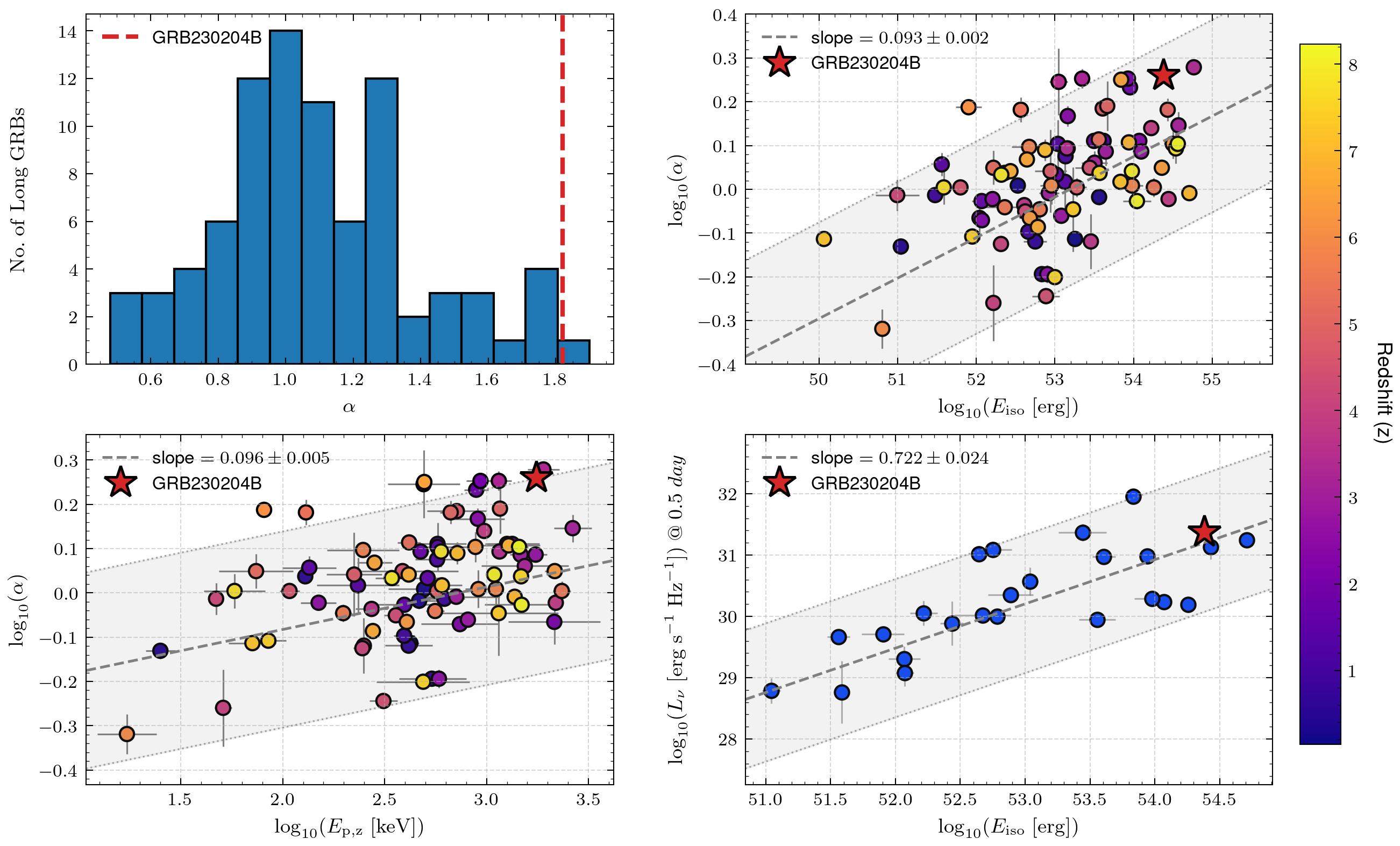}
    \caption{Top-left panel: histogram showing the distribution of temporal PL decay index ($\alpha$) in the optical afterglow for a sample of 86 LGRBs. \thisgrb, marked by a dashed red line, exhibits a very steep decay.
    Top-right panel: correlation between $\alpha$ and the isotropic equivalent prompt energy ($E_{\mathrm{iso}}$), showing a positive trend (dashed line) with \thisgrb\ (red star) located at the high-energy, steep decay end.
    Bottom-left panel: relation between $\alpha$ and the peak energy ($E_{\rm p,z}$) of the prompt emission, again highlighting \thisgrb\ as an outlier with high $E_{\rm p,z}$ and steep afterglow decay.
    Bottom-right panel: Afterglow optical luminosity ($L_\nu$) at 0.5 days (rest frame, $r$ band) as a function of $E_{\mathrm{iso}}$, illustrating the correlation between prompt energy release and afterglow brightness of a sample of 24 LGRBs.
    The shaded region shows the $95\%$ confidence level of the fits. Color bar encodes GRB redshift, indicating a broad redshift distribution across the sample.
    The data sample is provided in the supplementary material at \cite{swain_2025_17782846}, and the information has been compiled from multiple sources \citep{2001ApJ...561..178G, 2006ApJ...650..261S, 2007A-A...469L..13M, 2007AJ....133..122H, 2010ApJ...720.1513K, 2011A-A...526A.113F, 2011ApJ...743..154C, 2012ApJ...752L...6U, 2012MNRAS.421.1874G, 2012ApJ...748...59G, 2013ApJ...763...71A, 2013A-A...556A..23E, 2014ApJ...785...84J, 2014A-A...562A..29N, 2014MNRAS.440.1810M, 2014A-A...568A..19C, 2015A-A...577A..44O, Wang_2015, 2015ApJ...810...31V, 2016MNRAS.455..712L, 2016ApJ...817..152X, 2017ApJ...837...50G, 2017PASJ...69...20H, 2018ApJ...859..163H, 2018PASJ...70...92T, 2018A-A...620A.119D, 2019ApJ...872..118B, 2019A-A...632A.100H, 2020ApJ...894...43K, 2020ApJ...900..176L, 2021MNRAS.505.2662J, 2022A-A...658A..11M, 2022JCAP...06..034G, 2022A-A...665A.125R, 2022MNRAS.516.1584S, 2022ApJ...931...90L, 2022ApJ...929...16G, 2023ApJ...948...30Z, 2023MNRAS.523.4923P, 2023ApJ...959..118Z, 2023NatAs...7..843O, 2023ApJ...949L...4L, 2024ApJ...972..195L, 10.1093/mnras/stae1484, 2024MNRAS.527.8140A, 2025ApJ...978...29T}.
    }
    \label{fig:parameter_correlation}
\end{figure*}

In this section, we examine the correlation between the prompt and afterglow properties. The details of the samples collected for this analysis are presented in Appendix~\ref{appendix_corrsample}.

\vbdone Figure~\ref{fig:parameter_correlation} presents a comparative analysis of \thisgrb\ against a larger sample of LGRB. Notably, \thisgrb\ exhibits a steeper decay during the adiabatic phase ($\nu_m < \nu < \nu_c$) of the afterglow compared to most LGRBs. We find that $\alpha$ is correlated with isotropic equivalent energy and weakly correlated with the spectral peak energy. We also find that the $r$-band luminosity at 0.5 day after the burst is strongly correlated with the isotropic equivalent energy:

\begin{align}
\log_{10} \alpha &= (0.093 \pm 0.002)\, \log_{10}\left( \frac{E_{\mathrm{iso}}}{10^{51}~\mathrm{erg}} \right) \notag \\
&\quad - (4.93 \pm 0.13) \label{eq:alpha_eiso}
\end{align}

\begin{align}
\log_{10} \alpha &= (0.098 \pm 0.006)\, \log_{10}\left( \frac{E_{\rm p,z}}{1000~\mathrm{keV}} \right) \notag \\
&\quad - (0.28 \pm 0.01) \label{eq:alpha_ep}
\end{align}

\begin{align}
\log_{10} \left( \frac{L_r}{10^{30}~\mathrm{erg}} \right)
                &= (0.722 \pm 0.024)\, \log_{10}\left( \frac{E_{\mathrm{iso}}}{10^{51}~\mathrm{erg}} \right) \notag \\
                &\quad - (8.0 \pm 1.3). \label{eq:lr_eiso}
\end{align}
 
\vbdone These trends are broadly consistent with findings reported by \citet{2015MNRAS.453.4121O}. \thisgrb\ satisfies the observed correlation between the afterglow temporal decay index ($\alpha$) and the isotropic equivalent prompt energy ($E_{\mathrm{iso}}$) within the $95\%$ confidence region, as illustrated in Figure~\ref{fig:parameter_correlation}. Its relatively high $\alpha$ value is consistent with expectations for a wind-type CBM, which typically produces steeper afterglow decay. In the $\alpha$--$E_{\rm p,z}$ plane, \thisgrb\ lies near the boundary of the 2$\sigma$ confidence region, indicating that the $E_{\rm p,z}$ is lower than expected. On the other hand, in the $L_r$-$E_{\mathrm{iso}}$ relation, \thisgrb\ clearly satisfies the correlation, owing to both the bright optical afterglow and the energetic prompt emission of this GRB.

These correlations are not predicted by the standard fireball model, which treats prompt and afterglow phases as physically decoupled, with the former arising from internal dissipation within the jet and the latter from external shocks with the CBM \citep{1998ApJ...497L..17S, 2015PhR...561....1K}. However, if both phases originate from a common energy reservoir injected by the central engine, then more energetic GRBs could naturally produce both brighter prompt emission (a few percent of the total energy) and more luminous afterglows powered by the residual kinetic energy.

Furthermore, the brightness and temporal decay of GRB afterglows are governed by several key factors: (i) the density profile of the CBM, (ii) the energetics and internal structure of the jet, and (iii) geometric effects, such as the observer's viewing angle relative to the jet axis. The first two factors are closely related to the properties of the progenitor and the formation of the central engine. As a result, this leads to the expectation of correlations between the prompt emission and the afterglow observables.

In the collapsar framework, the mass-loss rate and metallicity of the progenitor star are crucial factors in determining whether a black hole forms and whether enough angular momentum is retained to generate a relativistic jet through an accretion disk \citep{Woosley_2006}. These parameters simultaneously shape the jet energetics and the CBM into which the jet propagates. In contrast, the viewing angle is an extrinsic parameter and primarily introduces scatter into these correlations.

Based on our analysis, \thisgrb\ is most plausibly associated with a W-R progenitor that generated a wind-type CBM while retaining sufficient core mass to collapse into a black hole capable of launching an energetic relativistic jet. This scenario naturally explains the highly luminous prompt emission, followed by a bright optical afterglow with a steep decay slope. The combined prompt and afterglow properties of \thisgrb\ are broadly consistent with the empirical correlations observed in our comparison sample of GRBs.

\section{Summary}
We present a comprehensive multiwavelength study of \thisgrb, which combines detailed prompt emission spectroscopy with broadband afterglow observations spanning X-rays to radio wavelengths. The prompt emission is characterized by multiple episodes and shows a significant thermal (photospheric) component. Our time-resolved spectral analysis indicates that a PL+BB model is statistically preferred over the empirical Band model for most temporal bins. This allows us to directly constrain the jet's physical properties, including the bulk Lorentz factor, photospheric radius, and base radius of the outflow. The inferred Lorentz factors are high ($\Gamma_j \sim 400$–600), suggesting a highly relativistic jet powered by an efficient central engine. The presence of a significant BB component disfavors a Poynting flux-dominated jet driven by magnetic reconnection and instead indicates a baryon-loaded outflow with weak to moderate magnetization.

The optical afterglow of \thisgrb\ shows a rapid, single PL decay with a temporal index of $\alpha \simeq 1.82$, while late-time radio observations reveal a long-lived plateau. Through closure relation analysis and semi-analytical modeling, we find that these characteristics can be naturally explained by a relativistic jet interacting with a wind-type CBM.  The afterglow modeling indicates a top-hat jet, with an electron PL index of $p = 2.76 \pm 0.01$, a wind density parameter of $A_\star \simeq 0.15$, and a low magnetic field fraction of $\epsilon_b \lesssim 10^{-4}$. The inferred isotropic equivalent kinetic energy is $E_{K,\mathrm{iso}} \gtrsim 2.7 \times 10^{54}$~erg, suggesting a moderately broad range of radiative efficiency that aligns with values typically associated with energetic LGRBs. The true energy, $E_{\gamma} \gtrsim 10^{52}$~erg, inferred from both prompt and afterglow analysis, places this GRB in the category of hyper-energetic GRBs.

We further examine \thisgrb\ within the framework of empirical correlations between prompt emissions and afterglow characteristics. Our findings indicate that the steep optical decay and high luminosity are consistent with the correlations among afterglow decay rates, optical luminosity, and prompt isotropic energy. This supports the idea of a physical connection between the prompt and afterglow phases, which is influenced by the central engine and the circumburst environment.

The combined properties of the prompt emission and afterglow of \thisgrb\ are best explained by a low-metallicity W-R progenitor that experienced moderate mass loss. This process created a wind-type CBM, while allowing the progenitor to retain enough angular momentum to form a black hole capable of launching an energetic relativistic jet. Overall, \thisgrb\ presents a compelling case where the prompt photospheric emission, rapidly fading optical afterglow, and late-time radio behavior can all be understood within a unified physical framework. This reinforces the connection between the properties of the progenitor, jet physics, and the observed phenomena associated with GRBs.

\begin{acknowledgments}
\added{We thank the anonymous referee for the comments and suggestions that have helped us to improve the paper.}

We thank all the people from the GROWTH collaboration for the observation.

GIT \citep{2022AJ....164...90K} is a 70~cm telescope with a $0^\circ.7$ field of view, set up by the Indian Institute of Astrophysics (IIA) and the Indian Institute of Technology Bombay (IITB) with funding from DST-SERB and IUSSTF. It is located at the Indian Astronomical Observatory (Hanle), operated by IIA. We acknowledge funding by the IITB alumni batch of 1994, which partially supports the operations of the telescope. Telescope technical details are available at \url{https://sites.google.com/view/growthindia/}.

The National Radio Astronomy Observatory and Green Bank Observatory are facilities of the U.S. National Science Foundation operated under cooperative agreement by Associated Universities, Inc. We thank the staff of the GMRT who made these observations possible. GMRT is run by the National Centre for Radio Astrophysics of the Tata Institute of Fundamental Research.

Based on observations obtained with the Samuel Oschin Telescope 48 inch and the 60 inch Telescope at the Palomar Observatory as part of the ZTF project. ZTF is supported by the National Science Foundation under Grant No. AST-2034437 and a collaboration including Caltech, IPAC, the Weizmann Institute of Science, the Oskar Klein Center at Stockholm University, the University of Maryland, Deutsches Elektronen-Synchrotron and Humboldt University, the TANGO Consortium of Taiwan, the University of Wisconsin at Milwaukee, Trinity College Dublin, Lawrence Livermore National Laboratories, IN2P3, University of Warwick, Ruhr University Bochum, Cornell University, and Northwestern University. Operations are conducted by COO, IPAC, and UW.

SED Machine is based upon work supported by the National Science Foundation under grant No. 1106171

We also acknowledge the Zenodo repository for sharing our data products.

\end{acknowledgments}





%
\facilities{\emph{MAXI} (GSC), \emph{Fermi} (GBM), \emph{Astrosat} (CZTI), \emph{Swift} (XRT), GIT, PO:1.5 (Palomar 60-inch telescope), GMRT, VLA}

\software{astropy \citep{2013A&A...558A..33A,2018AJ....156..123A,2022ApJ...935..167A}, Source Extractor \citep{1996A&AS..117..393B}, Astro-SCRAPPY \citep{2019ascl.soft07032M}, \sw{solve-field} astrometry engine \citep{2010AJ....139.1782L}, \sw{PSFEx} \citep{2013ascl.soft01001B}, ThreeML \citep{2015arXiv150708343V}, GDT-Fermi \citep{GDT-Fermi}}


\appendix 
\twocolumngrid 

\section{Data and Reduction}\label{appendixData}
For the prompt analysis, we primarily analyzed data from \emph{Fermi}-GBM and \emph{AstroSat}-CZTI. For broadband afterglow analysis, we reduced the data from observations by GIT, the Palomar 60 inch telescope (P60), and the Giant Metrewave Radio Telescope (GMRT), as described below.

\subsection{\emph{Fermi-GBM}}
\vbdone \emph{Fermi}-GBM was triggered at time $T_0 = \mathrm{21{:}47{:}51}$ UT on 2023 Feb 04 \citep{2023GCN.33288....1P}. Here, we report the results of an analysis of the \emph{Fermi}-GBM prompt data. GBM has 12 NaI detectors ($8~\mathrm{keV}$--$1~\mathrm{MeV}$) and two BGO detectors ($200~\mathrm{keV}$--$40~\mathrm{MeV}$). We choose the detectors according to the brightness of the spectral counts and the angle of the detector's boresight to the source location. The source was detected as bright in the NaI~8 detector, at $18^\circ$ away from the source. For analysis, we used two NaI detectors, NaI~7 and NaI~8, and a single BGO detector, BGO~1. The angle from the \emph{Fermi} LAT boresight at the GBM trigger time is $106^\circ$; thus, LAT data are not included in the analysis \citep{Ackermann_2012}. \added{The background-subtracted light curves are shown in Figure~\ref{fig:multi_panel}.} The duration of the GRB is $T_{90} = 216$~s in the energy range of $50$--$300~\mathrm{keV}$. The strongest peak occurred at $T_0 + 156~\mathrm{s}$ in the energy range of $10$--$1000~\mathrm{keV}$ with a photon flux of $7.3 \pm 0.4~\mathrm{ph~s^{-1}~cm^{-2}}$ \citep{2023GCN.33288....1P}.

\subsection{\emph{AstroSat CZTI}}
\vbdone The source was also seen by \emph{AstroSat}-CZTI \citep{2017JApA...38...31B}, where both the CZT modules ($20$--$200$~keV) and the Veto detectors ($100$--$500$~keV) detected multiple peaks of the emission~\citep{2021GCN.29410....1W}. The strongest peak in the CZT detectors occurred at $T_0 + 51.5$~s, whereas in the Veto detectors the strongest peak occurred at $T_0 + 50.8$~s. Our detailed reanalysis of the data shows that the peak count rate in the CZT was $548\pm50~\mathrm{counts~s}^{-1}$ above background, with a total of $23,613^{+1,157}_{-353}$ counts and a $T_{90}$ duration of $221.9^{+4.2}_{-5.3}$~s, consistent with the \emph{Fermi}-GBM $T_{90}$. In the Veto, the peak count rate was $1324^{+78}_{-86}~\mathrm{counts~s}^{-1}$ above background, with a total of $51,434^{+1332}_{-1566}$~counts and a $T_{90}$ of $229^{+22}_{-28}$~s. A total of \(3240\) Compton events were associated with the burst; however, the detector was at a boresight angle of $75^\circ$, which is too large to perform polarization analysis \citep{2022ApJ...936...12C}.

\subsection{GIT}\label{GIT}
\vbdone We used GIT located at the Indian Astronomical Observatory (IAO), Hanle-Ladakh, to acquire data of \thisgrb\ optical afterglow \citep{2022AJ....164...90K}. GIT is a 0.7~m, wide-field, fully robotic telescope specifically designed for the study of transient astrophysical events. The afterglow was observed in a Sloan~$r{^\prime}$ filter starting at 1.57~h after the GRB detection at 2025-02-04T21:47:51 UT by \emph{MAXI}-GSC \citep{2023GCN.33265....1S}. Data were downloaded and processed in real time by the GIT data reduction pipeline. Basic reduction, astrometry, and point-spread function (PSF) photometry followed the procedure described in \citet{2022AJ....164...90K}. The final magnitudes were calibrated against Pan-STARRS.

\subsection{P60/Spectral Energy Distribution Machine}
\vbdone Once the afterglow was identified, we used the Rainbow Camera on the Spectral Energy Distribution Machine (SEDM; \citealt{2018PASP..130c5003B, 2019A&A...627A.115R}) mounted on the P60 telescope to acquire $r$ and $i$-band imaging in 300~s exposures, starting 10.88~hr after the burst. The SEDM images were processed with a Python-based pipeline version of \texttt{Fpipe} \citep[Fremling Automated Pipeline;][]{FrSo2016}, which includes photometric calibrations. Afterglow spectroscopy was attempted, but unfortunately, the weather conditions did not allow for good signal-to-noise observations.

\subsection{Radio}
We conducted multiple follow-up observations with the Giant Metrewave Radio Telescope (GMRT) and the Very Large Array (VLA) at different epochs following the initial trigger. The GMRT observations, carried out on 2023 February 22, obtained an upper limit in Band 5 (1000--1450~MHz). The VLA observations, beginning on 2023 February 15, and extending over several epochs, yielded multiple detections in the X band (8--12~GHz; central frequency 10~GHz) and upper limits in the S band (2--4~GHz). A summary of the measured fluxes and upper limits from all observing epochs is presented in the supplementary material at \cite{swain_2025_17782846}\footnote{\url{https://zenodo.org/records/17782846/files/Table_afterglow-obs.csv}}.

\subsection{\emph{Swift-XRT}}
\vbdone Observations with \emph{Swift}-XRT began approximately $80.6~\mathrm{ks}$ after the trigger time. For the light curve analysis, we utilized the results publicly available from the UK Swift Science Data Centre (UKSSDC)\footnote{\url{https://www.swift.ac.uk/xrt_live_cat/00021535/}}. In the afterglow analysis, we considered two data points of unabsorbed flux density at $10~\mathrm{keV}$, and ignored the very weak upper limit at the third epoch.

\vbdone We reanalyzed the data of the first interval to obtain some spectral parameters. Since the optical light curve showed no evidence of flares or temporal breaks, we performed a time-integrated analysis over the interval from $T_0 + 80.6~\mathrm{ks}$ to $110.0~\mathrm{ks}$ in the $0.3$--$10~\mathrm{keV}$ energy band. The X-ray data were obtained from \emph{Swift} observations as described by \citet{eva07}. The total number of observed counts was 17, which is insufficient for spectral fitting due to the large number of free parameters. Therefore, we fixed the value of the intrinsic hydrogen column density to the mean value of $N_{\mathrm{H}} = 2.2 \times 10^{21}~\mathrm{cm^{-2}}$ as reported by \citet{eva07}. With this assumption, we performed a likelihood-based spectral fit using \texttt{ThreeML}, yielding a photon index of $1.4 \pm 0.4$. It corresponds to a spectral index of $\beta = 0.4 \pm 0.4$. The resulting flux is $2.3 \times 10^{-13}~\mathrm{erg~s^{-1}~cm^{-2}}$ in the $0.3$--$10~\mathrm{keV}$ band. The results are consistent with the automated analysis provided by the UKSSDC.

\section{Empirical Correlations of Prompt Emission}\label{appendix_prompt_corr}

\subsection{$K$-correction and Rest-frame Energetics}\label{kcorrection}
\vbdone To compare individual events in the broader GRB population, it is essential to calculate physical properties in a common rest-frame energy band, typically $1$--$10^{4}~\mathrm{keV}$. This requires transforming the observed quantities to the source rest frame using a $k$-correction factor, defined as
\begin{equation}\label{k_corr}
    k \equiv \frac{\int_{{1}/{(1+z)}}^{{10^4}/{(1+z)}} {E N(E)} dE}{\int_{e_1}^{e_2} E N(E) dE}, 
\end{equation}
where $N(E)$ is the spectral model in units of counts per unit energy, and $(e_1, e_2)$ brackets the energy band of the detector. 

\vbdone The isotropic equivalent energy in the rest frame, $E_\mathrm{iso}$, is calculated as
\begin{equation}\label{Eiso_eq}
E_{\mathrm{iso}} = \frac{4\pi D_L^2 \cdot k \cdot S_\gamma}{1+z},
\end{equation}
where $S_\gamma$ is the gamma-ray fluence, $D_L$ is the luminosity distance, $z$ is the redshift of the source, and $k$ is the $k$-correction factor \citep{Bloom_2001}. \added{The ``$1+z$" term in the denominator corrects for the cosmological time dilation of the burst duration, while the $k$-correction accounts for the spectral shift, effectively converting the observed fluence into the rest-frame energy (typically 1–10,000~keV).}

\subsection{Amati and Yonetoku Relations}\label{appendix_amati}
\begin{figure*}[th]
    \centering
    \begin{subfigure}[t]{0.48\textwidth}
        \includegraphics[width=\linewidth]{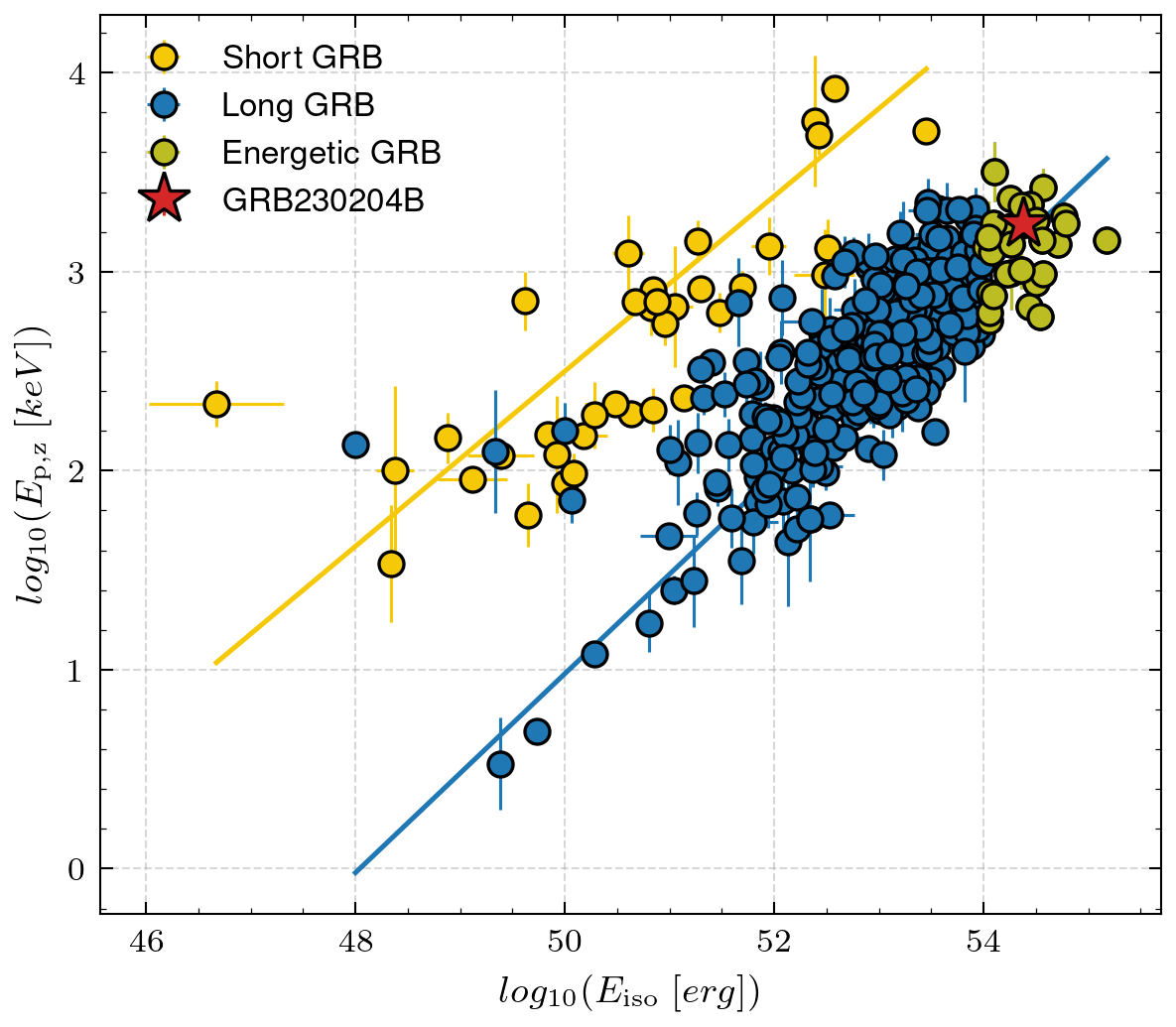}
        \caption{Amati relation: Correlation between rest-frame peak energy $E_{\rm p,z}$ and an isotropic equivalent energy $E_{\mathrm{iso}}$ for GRBs with known redshifts, adapted from \citet{2023ApJ...949L...4L}. \thisgrb\ is marked with a red star, consistent with the Amati relation and among the energetic long GRBs.}
        \label{fig:amati}
    \end{subfigure}%
    \hfill
    \begin{subfigure}[t]{0.48\textwidth}
        \includegraphics[width=\linewidth]{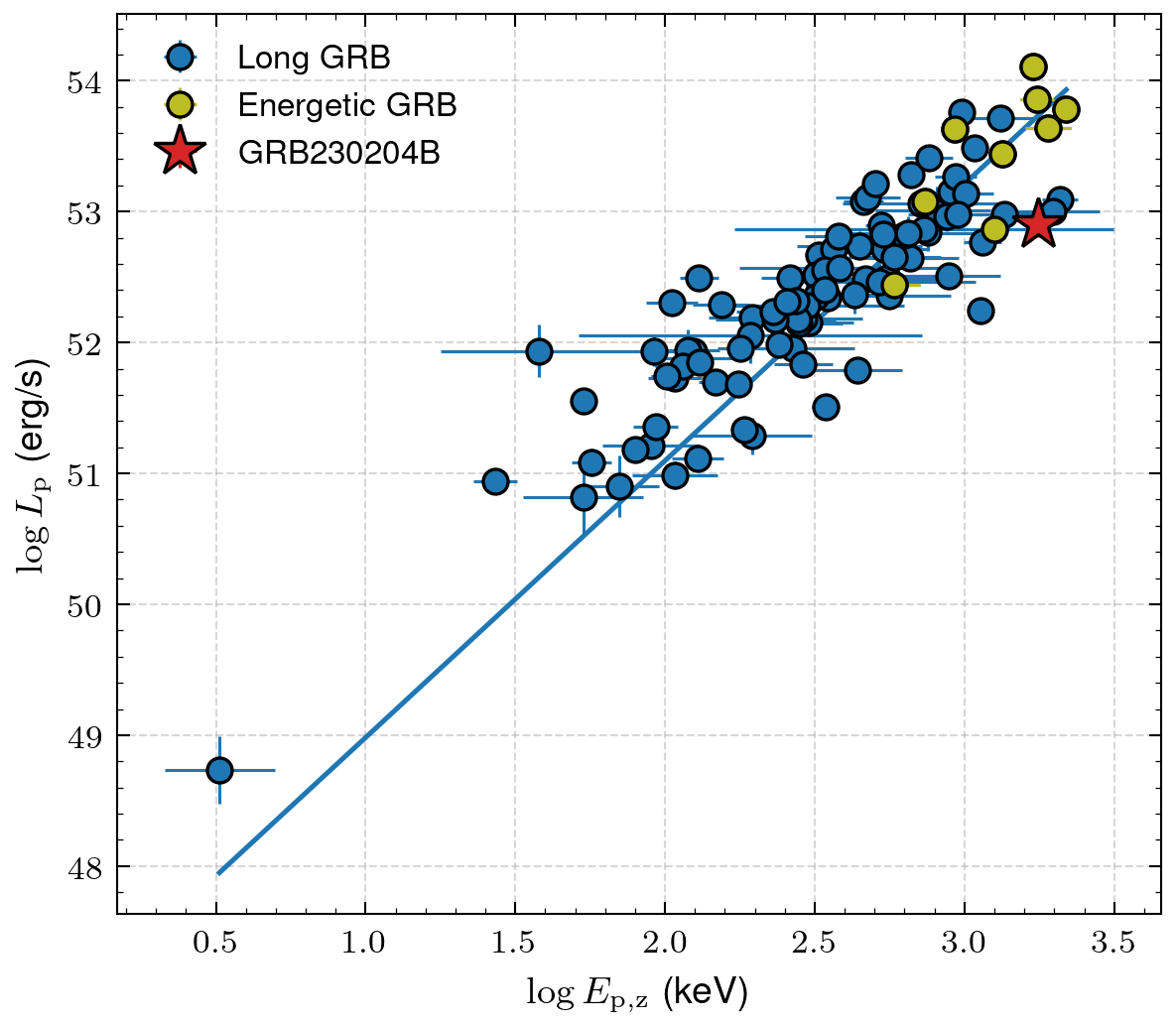}
        \caption{Yonetoku relation: Correlation between rest-frame peak energy $E_{\rm p,z}$ and peak luminosity $L_{\rm p}$, based on \citet{2010PASJ...62.1495Y}. \thisgrb\ (red star) lies within the energetic GRB population and follows the correlation.}
        \label{fig:Yonetoku}
    \end{subfigure}
    \caption{Correlation plots for \thisgrb: (a) Amati relation and (b) Yonetoku relation. Both panels show that \thisgrb\ belongs to the population of energetic long GRBs and conforms well to established empirical relations.}
    \label{fig:amati_yonetoku_combined}
\end{figure*}

We used the catalog from \cite{2023ApJ...949L...4L} to study this GRB in relation to the Amati relation \citep{2002A&A...390...81A}, which describes the correlation between the isotropic equivalent bolometric energy of the GRB and its rest-frame peak energy. This correlation shows a positive dependence, as illustrated in Figure~\ref{fig:amati}, and is typically expressed as follows:
\begin{equation}
    \frac{E_{\rm p,z}}{100~\mathrm{keV}} = C \times \left[\frac{E_{\mathrm{iso}}}{10^{52}~\mathrm{erg}}\right]^a,
\end{equation}
where $a$ lies between 0.4 and 0.6. 
This relationship suggests that energetic GRBs tend to release most of their energy in a harder spectrum. Notably, while \thisgrb\ follows this relation, its isotropic energy ($E_{\mathrm{iso}}$) is significantly higher than that of typical LGRBs, placing it in the category of highly energetic GRBs.

\vbdone We further explored the Yonetoku relation using the catalog given in \citet{2010PASJ...62.1495Y}. It is a tighter relation (Equation~\ref{eq:yonetoku}) between the peak energy in the rest frame and the peak luminosity of the GRB's prompt emission. The relation can be defined as
\begin{equation}
    \frac{L_p}{10^{52}~\mathrm{erg/s}} = C \times \left[\frac{E_{\rm p,z}}{1~\mathrm{keV}}\right]^b.\label{eq:yonetoku}
\end{equation}

Our analysis yields a best-fit slope of $b = 1.55 \pm 0.08$ and an intercept of $C = -3.52$, as illustrated in Figure~\ref{fig:Yonetoku}. The figure shows that the energetic GRBs exhibit higher values of both $E_{\rm p,z}$ and $L_{\rm p}$. However, they are not distinctly separated from the rest of the LGRBs. For \thisgrb, we obtain $L_{\rm p} = (8.7 \pm 2.5) \times 10^{52}~\mathrm{erg~s^{-1}}$ at $T_0+156$~s. This burst conforms well to the Yonetoku relation and aligns with the population of energetic GRBs.

\section{GRB Samples Used for Prompt Emission and Afterglow Properties Correlation Analysis}\label{appendix_corrsample}
To investigate the correlations between prompt emission and afterglow properties, we constructed a sample based on the availability of reliable prompt and optical afterglow measurements.

First, we utilized the catalog from \citet{2023ApJ...949L...4L} to obtain the redshift and prompt properties of the GRBs. To incorporate afterglow properties, we consider a collimated outflow without energy injection, i.e., $L_{\nu} \propto F_{\nu} \propto t^{-\alpha}\nu^{-\beta}$, where $\alpha$ and $\beta$ are linearly related by closure relations depending on the medium profile and the order of synchrotron frequencies ($\nu_a, \nu_m$ and $\nu_c$) \citep{1998ApJ...497L..17S, 2013NewAR..57..141G}.

\vbdone As discussed in \S\ref{sec:fast_fading}, most of the GRBs are observed in the adiabatic segment $\nu_m < \nu < \nu_{c}$, where the spectral index $\beta$ is expected to be $p/2$ and $\alpha$ depending on the profile of the medium. From the literature, we selected a sample of 86 LGRBs that have well-sampled optical afterglow light curves covering both early and late times, ensuring that the obtained $\alpha$ are consistent with the adiabatic segment. The derived temporal indices $\alpha$ for this sample are available in the supplementary material at \cite{swain_2025_17782846}\footnote{\url{https://zenodo.org/records/17782846/files/Table_grb-sample-correlation.csv}}.

\vbdone Next, we considered a subset of 24 LGRBs from this sample for which we could reliably calculate the optical luminosity in the $R$ or $ r$ band at 0.5 day postburst. We used extinction-corrected data from \citet{10.1093/mnras/stae1484} and selected light curves with high signal-to-noise ratios. We excluded early-time data if they showed a significantly steeper decay (indicative of a reverse shock), and also excluded late-time data if they showed a change in the decay rate (indicative of jet breaks). Each light curve was then fitted with a single PL model, $F_\nu \propto t^{-\alpha}$, to determine the temporal decay index $\alpha$. The corresponding spectral indices, defined as $F_\nu \propto \nu^{-\beta}$, were compiled from various publications, which are referenced in the caption of Figure~\ref{fig:parameter_correlation}. For sources where the $R$ band is obtained, we converted it to the $r$ band using the $\beta$ values. We then computed the $k$-corrected luminosity using
\begin{equation}
    L_{\nu} = 4\pi D_{L}^2 F_{\nu} (1+z)^{\beta -1},
\end{equation}
where $D_{L}$ is the luminosity distance, and $F_{\nu}$ is the observed or calculated flux in the $r$ band at 0.5 day.


\bibliography{sample7}{}
\bibliographystyle{aasjournal}




\end{document}